\documentclass[preprint,preprintnumbers,amsmath,amssymb]{revtex4}
\pdfoutput=1
\usepackage{epsfig}
\usepackage{graphicx}
\usepackage{dcolumn}
\usepackage{bm}
\usepackage{threeparttable}
\usepackage{xcolor}
\usepackage[utf8]{inputenc}
\usepackage{graphicx}
\usepackage{subfigure}

\def\beq{\begin{equation}}
\def\eeq{\end{equation}}
\def\eeqn{\end{equation}}
\newcommand\iden{\leavevmode\hbox{\small1\normalsize\kern-.33em1}}

\newcommand{\bea} {\begin{eqnarray}}
\newcommand{\eea} {\end{eqnarray}}

\def\tb {t_\beta}
\def\sb  {s_{\beta}}
\def\cb  {c_{\beta}}

\def\lam{\lambda}

\def\hm{{\hat m}}

\let\jnfont=\rm
\def\NPB#1 {{\jnfont Nucl.\ Phys.\ B }{\bf #1} }
\def\PLB#1 {{\jnfont Phys.\ Lett.\ B }{\bf #1} }
\def\EPJC#1 {{\jnfont Eur.\ Phys.\ Jour.\ C }{\bf #1} }
\def\PRD#1 {{\jnfont Phys.\ Rev.\ D }{\bf #1} }
\def\PRL#1 {{\jnfont Phys.\ Rev.\ Lett.\ }{\bf #1} }
\def\MPLA#1 {{\jnfont Mod.\ Phys.\ Lett.\ A }{\bf #1} }
\def\JPG#1 {{\jnfont J.\ Phys.\ G }{\bf #1} }
\def\CTP#1 {{\jnfont Commun.\ Theor.\ Phys.\ }{\bf #1} }
\def\JHEP#1 {{\jnfont JHEP \ }{\bf #1} }
\def\NPPS#1 {{\jnfont Nucl.\ Phys.\ Proc.\ Suppl.\ }{\bf #1} }
\def\CPC#1 {{\jnfont Comput.\ Phys.\ Commun.\ }{\bf #1} }
\def\CPL#1 {{\jnfont Chin.\ Phys.\ Lett. }{\bf #1} }
\def\APPB#1 {{\jnfont Acta\ Phys.\ Polon.\ B }{\bf #1} }

\def\lsim{\raise0.3ex\hbox{$<$\kern-0.75em\raise-1.1ex\hbox{$\sim$}}}
\def\gsim{\raise0.3ex\hbox{$>$\kern-0.75em\raise-1.1ex\hbox{$\sim$}}}
\def\PR#1 {{\jnfont Phys.\ Rept. }{\bf #1} }
\def\CHC#1 {{\jnfont Chin.\ Phys.\ C }{\bf #1} }
\def\NIMA#1 {{\jnfont Nucl.\ Instrum.\ Meth.\ A }{\bf #1} }
\def\JCAP#1 {{\jnfont JCAP \ }{\bf #1} }
\def\ASA#1 {{\jnfont Astron.\ Astrophys.\ A }{\bf #1} }  

\begin{document}

\title{\ \\[10mm] Dark matter, electroweak phase transition and gravitational wave in the type-II two-Higgs-doublet model with a singlet scalar field}

\author{Xiao-Fang Han$^{1}$, Lei Wang$^{1}$, Yang Zhang$^{2,3}$}
 \affiliation{$^1$ Department of Physics, Yantai University, Yantai
264005, P. R. China\\
$^2$ School of Physics, Zhengzhou University, ZhengZhou 450001, P. R. China\\
$^3$ ARC Centre of Excellence for Particle Physics at the Tera-scale, School of Physics and Astronomy, Monash University, Melbourne, Victoria 3800, Australia
}


\begin{abstract}
In the framework of type-II two-Higgs-doublet model with a singlet scalar dark matter $S$, we study the dark matter observables, 
the electroweak phase transition, and the gravitational wave signals by such strongly first order phase transition after imposing the constraints of the LHC Higgs data. We take the heavy CP-even Higgs $H$ as the only portal between the dark matter and
SM sectors, and find the LHC Higgs data and dark matter observables require $m_S$ and $m_H$ to be larger
 than 130 GeV and 360 GeV for $m_A=600$ GeV in the case of the 125 GeV Higgs with the SM-like coupling.
Next, we carve out some parameter space where a strongly first order electroweak phase transition can be achieved, and
find benchmark points for which the amplitudes of gravitational wave spectra 
reach the sensitivities of the future gravitational wave detectors.

\end{abstract}

\maketitle

\section{Introduction}
The weakly interacting massive particle is a primary candidate for dark matter
(DM) in the present Universe. Many extensions of SM have been proposed to provide a candidate of 
DM, and one simple extension is to add a singlet scalar DM to the type-II two-Higgs-doublet model (2HDM) \cite{2hdm,i-1,ii-2}.
The type-II 2HDM (2HDMIID) contains two CP-even states, $h$ and $H$, one neutral pseudoscalar $A$, 
 two charged scalars $H^{\pm}$, and one CP-even singlet scalar $S$ as the candidate of DM \cite{2hisos-0,2hisos-1,2hisos-2,2hisos-3,2hisos-4,2hisos-6,dmbu,1708.06882,1608.00421,1801.08317,1808.02667}.

In the type-II 2HDM model, the Yukawa couplings of the down-type quark and lepton can be both
enhanced by a factor of $\tan\beta$. Therefore, the flavor
observables and the LHC searches for Higgs can impose strong constraints on type-II 2HDM model.
In the 2HDMIID, the two CP-even states $h$ and $H$ may be portals between
the DM and SM sectors, and there are plentiful parameter space 
satisfying the direct and indirect experimental constraints of DM. 
The scalar potential of 2HDMIID contains the original one of type-II 2HDM and
one including DM.
For appropriate Higgs mass spectrum and coupling constants, the type-II 2HDM 
can trigger a strong first-order electroweak phase transition (SFOEWPT) 
in the early universe~\cite{PT_2HDM1, PT_2HDM1.5, PT_2HDM2, PT_2HDM3}, 
which is required by a successful explanation of the observed baryon asymmetry of the universe (BAU)~\cite{Sakharov:1967dj} 
and can produce primordial gravitational wave (GW) signals \cite{PT_GW}.

In this paper, we first examine the parameter space of the 2HDMIID using the recent LHC Higgs data and DM observables.
After imposing various theroretial and experimental constraints, we analyze whether a SFOEWPT is achievable in the 2HDMIID, and
discuss the resultant GW signals and its detectability at the future GW detectors,
such as LISA \cite{lisa}, Taiji \cite{taiji}, TianQin \cite{tianqin}, Big Bang Observer (BBO) \cite{bbodecigo},
DECi-hertz Interferometer GW Observatory (DECIGO) \cite{bbodecigo} and Ultimate-
DECIGO \cite{udecigo}.

Our work is organized as follows. In Sec. II we will give a brief introduction on the 2HDMIID. 
 In Sec. III and Sec. IV, we show the allowed parameter space
after imposing the limits of the LHC Higgs data and DM observables. In Sec. V, we examine the parameter space leading to a SFOEWPT
and the corresponding GW signal.
Finally, we give our conclusion in Sec. VI.

\section{Type-II two-Higgs-doublet model with a scalar dark matter}
The scalar potential of 2HDMIID is given as \cite{2h-poten}
\begin{eqnarray} \label{V2HDM} \mathcal{V}_{tree} &=& m_{11}^2
(\Phi_1^{\dagger} \Phi_1) + m_{22}^2 (\Phi_2^{\dagger}
\Phi_2) - \left[m_{12}^2 (\Phi_1^{\dagger} \Phi_2 + \rm h.c.)\right]\nonumber \\
&&+ \frac{\lambda_1}{2}  (\Phi_1^{\dagger} \Phi_1)^2 +
\frac{\lambda_2}{2} (\Phi_2^{\dagger} \Phi_2)^2 + \lambda_3
(\Phi_1^{\dagger} \Phi_1)(\Phi_2^{\dagger} \Phi_2) + \lambda_4
(\Phi_1^{\dagger}
\Phi_2)(\Phi_2^{\dagger} \Phi_1) \nonumber \\
&&+ \left[\frac{\lambda_5}{2} (\Phi_1^{\dagger} \Phi_2)^2 + \rm
h.c.\right]\nonumber\\
&&+{1\over 2}S^2(\kappa_{1}\Phi_1^\dagger \Phi_1
+\kappa_{2}\Phi_2^\dagger \Phi_2)+{m_{0}^2\over
2}S^2+{\lambda_S\over 4!}S^4.
\end{eqnarray}
Here we discuss the CP-conserving model in which all $\lambda_i$, $\kappa_i$ and
$m_{12}^2$ are real. The $S$ is a real singlet scalar field, and $\Phi_1$ and $\Phi_2$ are complex Higgs doublets with hypercharge $Y = 1$:
\begin{equation}
\Phi_1=\left(\begin{array}{c} \phi_1^+ \\
\frac{1}{\sqrt{2}}\,(v_1+\phi_1^0+ia_1)
\end{array}\right)\,, \ \ \
\Phi_2=\left(\begin{array}{c} \phi_2^+ \\
\frac{1}{\sqrt{2}}\,(v_2+\phi_2^0+ia_2)
\end{array}\right).
\end{equation}
Where $v_1$ and $v_2$ are the electroweak vacuum expectation values
(VEVs) with $v^2 = v^2_1 + v^2_2 = (246~\rm GeV)^2$, and the ratio of the two VEVs is defined
as $\tan\beta=v_2 /v_1$. 
The linear and cubic terms of the $S$ field are
forbidden by a $Z'_2$ symmetry, under which $S\rightarrow -S$. The $S$ is a possible DM candidate since
it does not acquire a VEV. After spontaneous electroweak symmetry breaking, the remaining physical states are three neutral
CP-even states $h$, $H$, and $S$, one neutral pseudoscalar $A$, and two charged
scalars $H^{\pm}$.

 We can obtain the DM mass and
the cubic interactions with the neutral Higgses from Eq.
(\ref{V2HDM}),
\begin{eqnarray}
m_S^2&=&m_0^2+\frac{1}{2}\kappa_1
v^2\cos^2\beta+\frac{1}{2}\kappa_2 v^2\sin^2\beta,\nonumber\\
-\lambda_{h} vS^2h/2&\equiv& -(-\kappa_{1}\sin\alpha\cos\beta+\kappa_{2}\cos\alpha\sin\beta)vS^2h/2,\nonumber\\
-\lambda_{H} vS^2H/2&\equiv&
-(\kappa_{1}\cos\alpha\cos\beta+\kappa_{2}\sin\alpha\sin\beta)vS^2H/2,
\label{dmcoup}\end{eqnarray}
with $\alpha$ being the mixing angle of $h$ and $H$.

The Yukawa interactions are written as
 \bea
- {\cal L} &=&Y_{u2}\,\overline{Q}_L \, \tilde{{ \Phi}}_2 \,u_R
+\,Y_{d1}\,
\overline{Q}_L\,{\Phi}_1 \, d_R\, + \, Y_{\ell 1}\,\overline{L}_L \, {\Phi}_1\,e_R+\, \mbox{h.c.}\,, \eea where
$Q_L^T=(u_L\,,d_L)$, $L_L^T=(\nu_L\,,l_L)$,
$\widetilde\Phi_{1,2}=i\tau_2 \Phi_{1,2}^*$, and $Y_{u2}$,
$Y_{d1}$ and $Y_{\ell 1}$ are $3 \times 3$ matrices in family
space.

The Yukawa couplings of the neutral Higgs bosons normalized to the SM are given by
\bea\label{hffcoupling} &&
y_{h}^{f_i}=\left[\sin(\beta-\alpha)+\cos(\beta-\alpha)\kappa_f\right], \nonumber\\
&&y_{H}^{f_i}=\left[\cos(\beta-\alpha)-\sin(\beta-\alpha)\kappa_f\right], \nonumber\\
&&y_{A}^{f_i}=-i\kappa_f~{\rm (for~u)},~~~~y_{A}^{f_i}=i \kappa_f~{\rm (for~d,~\ell)},\nonumber\\ 
&&{\rm with}~\kappa_d=\kappa_\ell\equiv-\tan\beta,~~~\kappa_u\equiv 1/\tan\beta.\eea 

The charged Higgs has the following Yukawa interactions,
\begin{align} \label{eq:Yukawa2}
 \mathcal{L}_Y & = - \frac{\sqrt{2}}{v}\, H^+\, \Big\{\bar{u}_i \left[\kappa_d\,(V_{CKM})_{ij}~ m_{dj} P_R
 - \kappa_u\,m_{ui}~ (V_{CKM})_{ij} ~P_L\right] d_j + \kappa_\ell\,\bar{\nu} m_\ell P_R \ell
 \Big\}+h.c.,
 \end{align}
where $i,j=1,2,3$.

The neutral Higgs boson couplings with the gauge bosons normalized to the
SM are given by
\beq
y^{V}_h=\sin(\beta-\alpha),~~~
y^{V}_H=\cos(\beta-\alpha),\label{hvvcoupling}
\eeq
where $V$ denotes $Z$ or $W$.
In the type-II 2HDM, the 125 GeV Higgs is allowed to
have the SM-like coupling and wrong sign Yukawa coupling, 
\bea
&&y_h^{f_i}~\times~y^{V}_h > 0~{\rm for~SM-like~coupling},~~~\nonumber\\
&&y_h^{f_i}~\times~y^{V}_h < 0~{\rm for~wrong~sign~Yukawa~coupling}.\label{wrongsign}
\eea

\section{The experimental constraints of the Higgs data at the LHC}
\subsection{Numerical calculations}
We take the light CP-even Higgs boson $h$ as the
SM-like Higgs, $m_h=125$ GeV. The measurement of the branching fraction of $b \to s\gamma$ gives the
stringent constraints on the charged Higgs mass of the type-II 2HDM, $m_{H^{\pm}} > 570$ GeV \cite{bsr570}.
If the 125 GeV Higgs boson is the portal between the DM and SM sectors, it is favored to
have wrong sign Yukawa coupling which can realize the isospin-violating DM interactions with nucleons and 
relax the bounds of direct detection of DM. However, Ref. \cite{PT_2HDM2} shows the the wrong sign Yukwa coupling region
of type-II 2HDM is strongly restricted by the requirement of SFOEWPT. Therefore, in this paper we take the heavy CP-even Higgs
$H$ as the only portal between the DM and SM sectors, and focus on the case of the 125 GeV with the SM-like couping.
The $S$, $T$, and $U$ oblique parameters give the stringent
constraints on the mass spectrum of Higgses of type-II 2HDM \cite{1604.01406,1701.02678,2003.06170}. 
One of $m_A$ and $m_H$ is around 600 GeV, and another is allowed to have 
a wide mass range including low mass \cite{1701.02678,2003.06170}. Therefore, we fix $m_A=600$ GeV to make the portal $H$ to have a 
wide mass range.

In our calculation, we consider the following observables and constraints:

\begin{itemize}
\item[(1)] Theoretical constraints. The scalar potential of the model contains one of the type-II 2HDM and one of the DM sector. 
The vacuum stability, perturbativity, and tree-level unitarity impose constraints on the relevant parameters,
which are discussed in detail in Refs. \cite{2hisos-4,2hisos-6}. Here we employ the formulas in \cite{2hisos-4,2hisos-6} to
implement the theoretical constraints. Compared to Refs. \cite{2hisos-4,2hisos-6}, there are additional factors of $\frac{1}{2}$
in the $\kappa_1$ term and the $\kappa_2$ term of this paper. In addition, we require that the potential has a global minimum at the point of ($<h_1>=v_1$, $<h_2>=v_2$, $<S_1>=0$).

\item[(2)] The oblique parameters. The $S$, $T$, $U$ parameters can impose stringent constraints on 
the mass spectrum of Higgses of 2HDM. We use $\textsf{2HDMC}$ \cite{2hc-1} to calculate the $S$, $T$, $U$ parameters.
Taking the recent fit results of Ref. \cite{pdg2018}, we use the following 
values of $S$, $T$, $U$,
\beq
S=0.02\pm 0.10,~~  T=0.07\pm 0.12,~~ U=0.00 \pm 0.09. 
\eeq
The correlation coefficients are 
\beq
\rho_{ST} = 0.89,~~  \rho_{SU} = -0.54,~~  \rho_{TU} = -0.83.
\eeq

\item[(3)] The flavor observables and $R_b$. We employ $\textsf{SuperIso-3.4}$ \cite{spriso} to 
calculate $Br(B\to X_s\gamma)$, and $\Delta m_{B_s}$ is calculated following the
formulas in \cite{deltmq}. Besides, we include the constraints of bottom quarks produced in $Z$ decays, $R_b$,
which is calculated following the formulas in \cite{rb1,rb2}.

\begin{table}
\begin{footnotesize}
\begin{tabular}{| c | c | c | c |}
\hline
\textbf{Channel} & \textbf{Experiment} & \textbf{Mass range [GeV]}  &  \textbf{Luminosity} \\
\hline
 {$gg/b\bar{b}\to H/A \to \tau^{+}\tau^{-}$} & ATLAS 8 TeV~\cite{47Aad:2014vgg} & 90-1000 & 19.5-20.3 fb$^{-1}$ \\
{$gg/b\bar{b}\to H/A \to \tau^{+}\tau^{-}$} & CMS 8 TeV~\cite{48CMS:2015mca} &  90-1000  &19.7 fb$^{-1}$ \\
{$gg/b\bar{b}\to H/A \to \tau^{+}\tau^{-}$} & CMS 13 TeV~\cite{add-hig-16-037} & 90-3200 &12.9 fb$^{-1}$ \\
{$gg/b\bar{b}\to H/A \to \tau^{+}\tau^{-}$} & CMS 13 TeV \cite{1709.07242}& 200-2250   & 36.1 fb$^{-1}$ \\
{$b\bar{b}\to H/A \to \tau^{+}\tau^{-}$} & CMS 8 TeV \cite{1511.03610}& 25-80   & 19.7 fb$^{-1}$ \\
{$gg/b\bar{b}\to H/A \to \tau^{+}\tau^{-}$} & ATLAS 13 TeV \cite{2002.12223}& 200-2500   & 139 fb$^{-1}$ \\
\hline
 {$b\bar{b}\to H/A \to \mu^{+}\mu^{-}$} & CMS 8 TeV~\cite{CMS-HIG-15-009} & 25-60 & 19.7 fb$^{-1}$ \\
\hline
 {$pp\to H/A \to \gamma\gamma$} & ATLAS 13 TeV \cite{80lenzi} & 200-2400 & 15.4 fb$^{-1}$ \\
{$gg\to H/A \to \gamma\gamma$}& CMS 8+13 TeV \cite{81rovelli}& 500-4000 & 12.9 fb$^{-1}$ \\
{$gg\to H/A \to \gamma\gamma$~+~$t\bar{t}H/A~(H/A\to \gamma\gamma)$}& CMS 8 TeV \cite{HIG-17-013-pas}& 80-110 & 19.7 fb$^{-1}$ \\
{$gg\to H/A \to \gamma\gamma$~+~$t\bar{t}H/A~(H/A\to \gamma\gamma)$}& CMS 13 TeV \cite{HIG-17-013-pas}& 70-110 & 35.9 fb$^{-1}$ \\
{$VV\to H \to \gamma\gamma$~+~$VH~(H\to \gamma\gamma)$}& CMS 8 TeV \cite{HIG-17-013-pas}& 80-110 & 19.7 fb$^{-1}$ \\
{$VV\to H \to \gamma\gamma$~+~$VH~(H\to \gamma\gamma)$}& CMS 13 TeV \cite{HIG-17-013-pas}& 70-110 & 35.9 fb$^{-1}$ \\
\hline

 {$gg/VV\to H\to W^{+}W^{-}$} & ATLAS 8 TeV  \cite{55Aad:2015agg}& 300-1500  &  20.3 fb$^{-1}$\\

{$gg/VV\to H\to W^{+}W^{-}~(\ell\nu\ell\nu)$} & ATLAS 13 TeV  \cite{77atlasww13}& 300-3000  &  13.2 fb$^{-1}$\\

{$gg\to H\to W^{+}W^{-}~(\ell\nu qq)$} & ATLAS 13 TeV  \cite{78atlasww13lvqq}& 500-3000  &  13.2 fb$^{-1}$\\

{$gg/VV\to H\to W^{+}W^{-}~(\ell\nu qq)$} & ATLAS 13 TeV  \cite{1710.07235}& 200-3000  &  36.1 fb$^{-1}$\\
{$gg/VV\to H\to W^{+}W^{-}~(e\nu \mu\nu)$} & ATLAS 13 TeV  \cite{1710.01123}& 200-3000  &  36.1 fb$^{-1}$\\

{$gg/VV\to H\to W^{+}W^{-}$} & CMS 13 TeV  \cite{1912.01594}& 200-3000  &  35.9 fb$^{-1}$\\
\hline

$gg/VV\to H\to ZZ$ & ATLAS 8 TeV \cite{57Aad:2015kna}& 160-1000 & 20.3 fb$^{-1}$ \\

$gg\to H \to ZZ(\ell \ell \nu \nu)$ & ATLAS 13 TeV~\cite{74koeneke4} & 300-1000  & 13.3 fb$^{-1}$ \\
$gg\to H\to ZZ(\nu \nu qq)$ & ATLAS 13 TeV~\cite{75koeneke5} & 300-3000 & 13.2 fb$^{-1}$ \\
$gg/VV\to H\to ZZ(\ell \ell qq)$ & ATLAS 13 TeV~\cite{75koeneke5} & 300-3000 & 13.2 fb$^{-1}$ \\
$gg/VV\to H\to ZZ(\ell\ell\ell\ell)$ & ATLAS 13 TeV~\cite{76koeneke3} & 200-3000 & 14.8 fb$^{-1}$ \\

$gg/VV\to H\to ZZ(\ell\ell\ell\ell+\ell\ell\nu\nu)$ & ATLAS 13 TeV~\cite{1712.06386} & 200-2000 & 36.1 fb$^{-1}$ \\
$gg/VV\to H\to ZZ(\nu\nu qq+\ell\ell qq)$ & ATLAS 13 TeV~\cite{1708.09638} & 300-5000 & 36.1 fb$^{-1}$ \\
\hline

\end{tabular}
\end{footnotesize}
\caption{The upper limits at 95\%  C.L. on the production cross-section times branching ratio of
$\tau^+\tau^-$, $\mu^+\mu^-$, $\gamma\gamma$, $WW$, and $ZZ$ considered in 
the $H$ and $ A $ searches at the LHC.}
\label{tabh}
\end{table}

\begin{table}
\begin{footnotesize}
\begin{tabular}{| c | c | c | c |}
\hline
\textbf{Channel} & \textbf{Experiment} & \textbf{Mass range [GeV]}  &  \textbf{Luminosity} \\
\hline
$gg\to H\to hh \to (\gamma \gamma) (b \bar{b})$ & CMS 8 TeV \cite{64Khachatryan:2016sey} & 250-1100  & 19.7 fb$^{-1}$\\

$gg\to H\to hh \to (b\bar{b}) (b\bar{b})$ & CMS 8 TeV \cite{65Khachatryan:2015yea}&   270-1100   & 17.9 fb$^{-1}$\\

$gg\to H\to hh \to (b\bar{b}) (\tau^{+}\tau^{-})$ & CMS 8 TeV \cite{66Khachatryan:2015tha}&  260-350 & 19.7 fb$^{-1}$\\

$gg \to H\to hh \to b\bar{b}b\bar{b}$ & ATLAS 13 TeV~\cite{84varol} & 300-3000  &  13.3 fb$^{-1}$ \\

$gg \to H\to hh \to b\bar{b}b\bar{b}$ & CMS 13 TeV~\cite{1710.04960} & 750-3000  &  35.9 fb$^{-1}$ \\
$gg \to H\to hh \to (b\bar{b}) (\tau^{+}\tau^{-})$ & CMS 13 TeV~\cite{1707.02909} & 250-900  &  35.9 fb$^{-1}$ \\

$pp \to H\to hh $ & CMS 13 TeV~\cite{1811.09689} & 250-3000  &  35.9 fb$^{-1}$ \\

$gg \to H\to hh \to b\bar{b}ZZ$ & CMS 13 TeV~\cite{2006.06391} & 260-1000  &  35.9 fb$^{-1}$ \\

$gg \to H\to hh \to b\bar{b}\tau^{+}\tau^{-}$ & CMS 13 TeV~\cite{2007.14811} & 1000-3000  &  139 fb$^{-1}$ \\
\hline

$gg\to A\to hZ \to (\tau^{+}\tau^{-}) (\ell \ell)$ & CMS 8 TeV \cite{66Khachatryan:2015tha}& 220-350 & 19.7 fb$^{-1}$\\

$gg\to A\to hZ \to (b\bar{b}) (\ell \ell)$ & CMS 8 TeV \cite{67Khachatryan:2015lba} & 225-600 &19.7 fb$^{-1}$ \\

$gg\to A\to hZ\to (\tau^{+}\tau^{-}) Z$ & ATLAS 8 TeV \cite{68Aad:2015wra}&220-1000 & 20.3 fb$^{-1}$ \\

 {$gg\to A\to hZ\to (b\bar{b})Z$} & ATLAS 8 TeV \cite{68Aad:2015wra}& 220-1000 & 20.3 fb$^{-1}$  \\

{$gg/b\bar{b}\to A\to hZ\to (b\bar{b})Z$}& ATLAS 13 TeV \cite{1712.06518}& 200-2000 & 36.1 fb$^{-1}$  \\

{$gg/b\bar{b}\to A\to hZ\to (b\bar{b})Z$}& CMS 13 TeV \cite{1903.00941}& 225-1000 & 35.9 fb$^{-1}$  \\

{$gg\to A\to hZ\to (\tau^{+}\tau^{-}) (\ell \ell)$}& CMS 13 TeV \cite{1910.11634}& 220-400 & 35.9 fb$^{-1}$  \\
\hline

 {$gg\to h \to AA \to \tau^{+}\tau^{-}\tau^{+}\tau^{-}$} & ATLAS 8 TeV~\cite{1505.01609} & 4-50 & 20.3 fb$^{-1}$ \\
{$pp\to  h \to AA \to \tau^{+}\tau^{-}\tau^{+}\tau^{-}$} & CMS 8 TeV~\cite{1701.02032} &  5-15  &19.7 fb$^{-1}$ \\
{$pp\to  h \to AA \to (\mu^{+}\mu^{-})(b\bar{b})$} & CMS 8 TeV~\cite{1701.02032} &  25-62.5  &19.7 fb$^{-1}$ \\
{$pp\to  h \to AA \to (\mu^{+}\mu^{-})(\tau^{+}\tau^{-})$} & CMS 8 TeV~\cite{1701.02032} &  15-62.5  &19.7 fb$^{-1}$ \\

{$pp\to  h \to AA \to (b\bar{b})(\tau^{+}\tau^{-})$} & CMS 13 TeV~\cite{1805.10191} &  15-60  &35.9 fb$^{-1}$ \\
{$pp\to  h \to AA \to \tau^{+}\tau^{-}\tau^{+}\tau^{-}$} & CMS 13 TeV~\cite{1907.07235} &  4-15  &35.9 fb$^{-1}$ \\
{$pp\to  h \to AA \to \mu^{+}\mu^{-}\tau^{+}\tau^{-}$} & CMS 13 TeV~\cite{2005.08694} &  3.6-21  &35.9 fb$^{-1}$ \\
\hline
$gg\to A(H)\to H(A)Z\to (b\bar{b}) (\ell \ell)$ & CMS 8 TeV \cite{160302991} & 40-1000 &19.8 fb$^{-1}$ \\

$gg\to A(H)\to H(A)Z\to (\tau^{+}\tau^{-}) (\ell \ell)$ & CMS 8 TeV \cite{160302991}& 20-1000 & 19.8 fb$^{-1}$ \\

$gg/b\bar{b}\to A(H)\to H(A)Z\to (b\bar{b}) (\ell \ell)$ & ATLAS 13 TeV \cite{1804.01126}& 130-800 & 36.1 fb$^{-1}$ \\

$gg\to A(H)\to H(A)Z\to (b\bar{b}) (\ell \ell)$ & CMS 13 TeV \cite{1911.03781}& 30-1000 & 35.9 fb$^{-1}$ \\
\hline
\end{tabular}
\end{footnotesize}
\caption{The upper limits at 95\%  C.L. on the production cross-section times branching ratio for the channels
of Higgs-pair and a Higgs production in association with $Z$ at the LHC.}
\label{tabhh}
\end{table}

\item[(4)] The global fit to the 125 GeV Higgs signal data. The version 2.0 of $\textsf{Lilith}$ \cite{lilith} is used to perform the
$\chi^2$ calculation for the signal strengths of the 125 GeV Higgs combining the LHC run-I and
run-II data (up to datasets of 36 fb$^{-1}$). We pay
particular attention to the surviving samples with
$\chi^2-\chi^2_{\rm min} \leq 6.18$, where $\chi^2_{\rm min}$
denotes the minimum of $\chi^2$. These samples correspond to be within
the $2\sigma$ range in any two-dimension plane of the
model parameters when explaining the Higgs data.

\item[(5)] The exclusion limits of searches for additional Higgs bosons. We use
$\textsf{HiggsBounds-4.3.1}$ \cite{hb1,hb2} to implement the exclusion
constraints from the neutral and charged Higgs searches at LEP at 95\% confidence level.

Because the $b$-quark
loop and top quark loop have destructive interference contributions to $gg\to A$ production in the type-II 2HDM, 
the cross section decreases with an increase
of $\tan\beta$, reaches the minimum value for the moderate $\tan\beta$, and is dominated by the $b$-quark loop for enough large
 $\tan\beta$. In addition to $\tan\beta$
and $m_H$, the cross section of $gg\to H$ depends on $\sin(\beta-\alpha)$. We employ $\textsf{SusHi}$ to 
compute the cross sections for $H$ and $A$ in the
gluon fusion and $b\bar{b}$-associated production at NNLO in QCD \cite{sushi}. 
The cross sections of $H$ via vector boson fusion process are deduced from results of the LHC
Higgs Cross Section Working Group \cite{higgswg}. We employ
$\textsf{2HDMC}$ to calculate the branching ratios of the
various decay modes of $H$ and $A$. The searches for the additional Higgs considered by us are listed in 
Tables \ref{tabh} and \ref{tabhh}.
The LHC searches for $H^\pm$ can not impose any constraints on the
model for $m_{H^{\pm}}>500$ GeV and 1 $\leq \tan\beta \leq 25$ \cite{mhp500}. Therefore, we do not
consider the constraints from the searches for the heavy charged Higgs. 
\end{itemize}

\subsection{Results and discussions}
\begin{figure}[tb]
 \epsfig{file=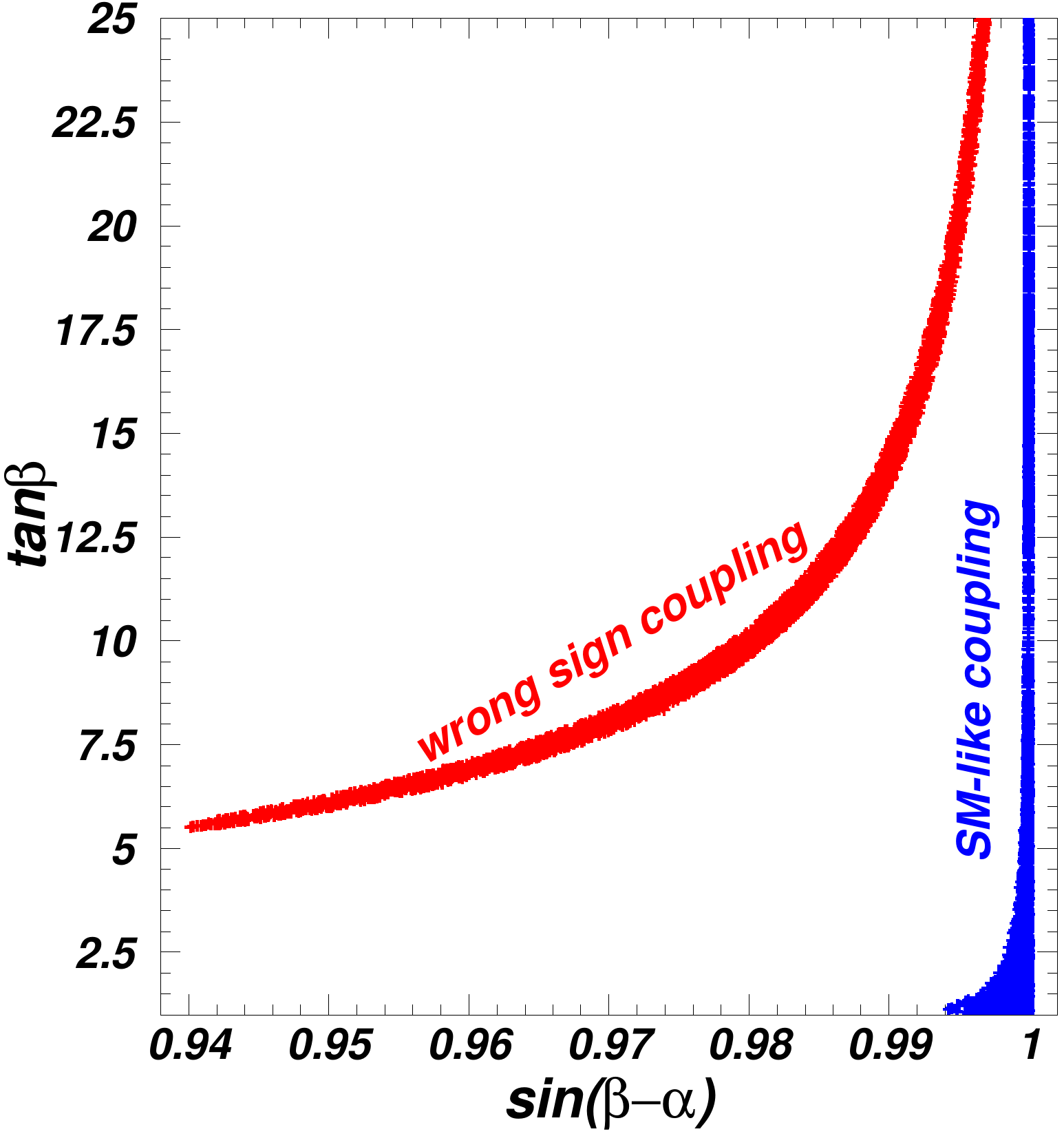,height=7.0cm}
 \epsfig{file=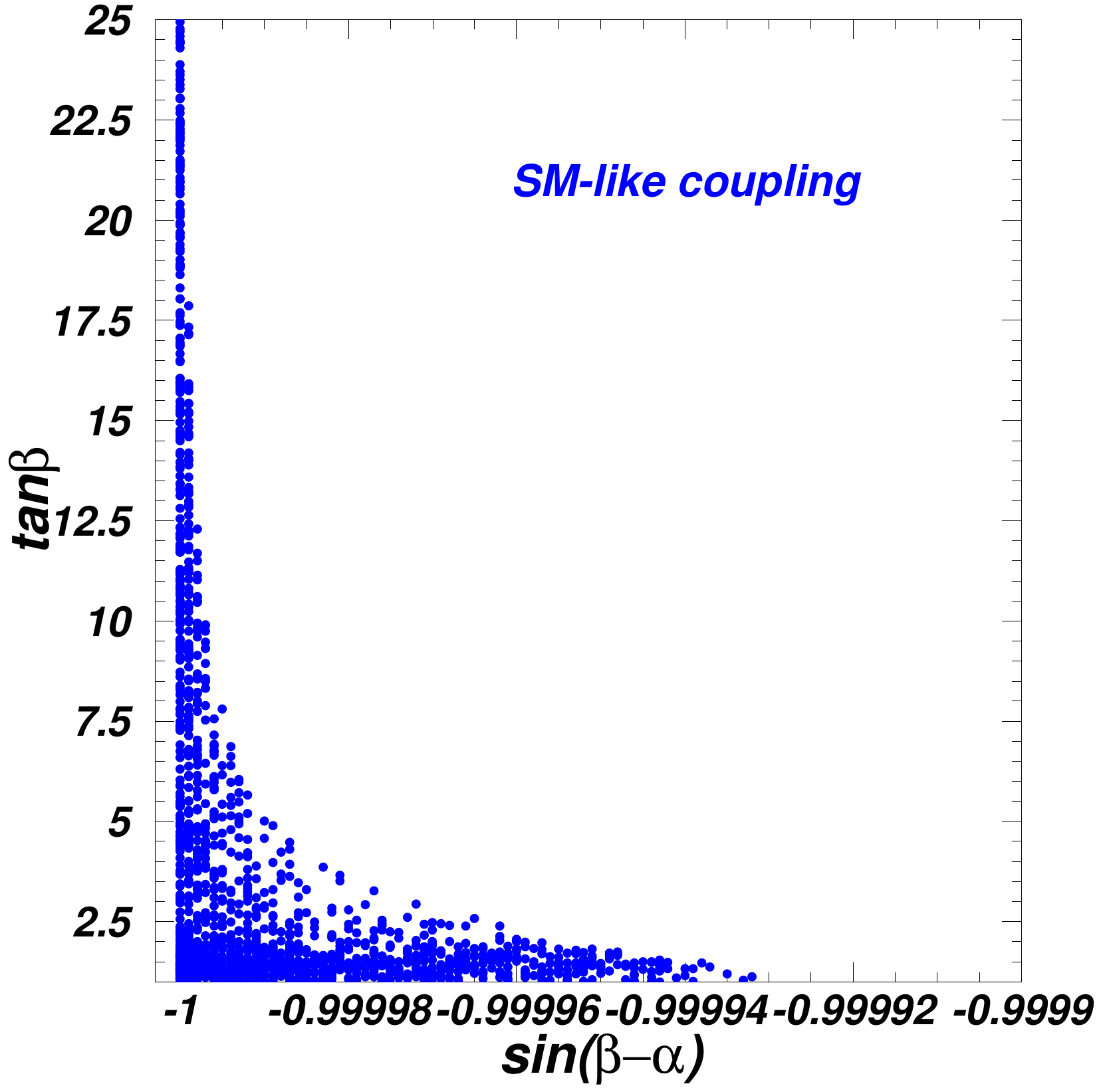,height=7.0cm}
\vspace{-0.0cm} \caption{Scatter plots of $\sin(\beta-\alpha)$ and $\tan\beta$ satisfying the constraints of the 125 GeV Higgs signal data.} \label{125higgs}
\end{figure}

In Fig. \ref{125higgs}, we show $\sin(\beta-\alpha)$ and $\tan\beta$ allowed by the 125 GeV Higgs signal data at the LHC. 
From Fig. \ref{125higgs}, we see that $\tan\beta$ and $\sin(\beta-\alpha)$ have strong correlation due to the constraints of the 125 GeV Higgs data,
especially for the case of the wrong sign Yukawa coupling. The wrong sign Yukawa coupling can be achieved only for $\sin(\beta-\alpha) > 0$,
and $\tan\beta$ is restricted to be in a very narrow range for a given $\sin(\beta-\alpha)$. 
For the case of the SM-like coupling, $\sin(\beta-\alpha)$ is required to be in two very narrow ranges of $-1.0\sim-0.99993$ and $0.994\sim1.0$.
The $\tan\beta$ is allowed to be as low as 1.0, and its upper bound increases with $\mid\sin(\beta-\alpha)\mid$ in the case of the
 the SM-like Higgs coupling.

Now we examine the parameter space of 2HDMIID using the exclusion limits of
searches for additional Higgses at the LHC. In the 2HDMIID, we take the heavy CP-even
Higgs $H$ as only portal between DM and SM sectors, and the decay $H\to SS$ opens for
$2m_S<m_H$. The decay mode possibly affects the allowed parameter space, but the constraints of the DM observables 
have to be simultaneously considered. Here we temporarily assume $2m_S>m_H$, and close the $H\to SS$ decay mode.
In the next section, the effects of $H\to SS$ will be considered by combining the DM observables.

\begin{figure}[tb]
 \epsfig{file=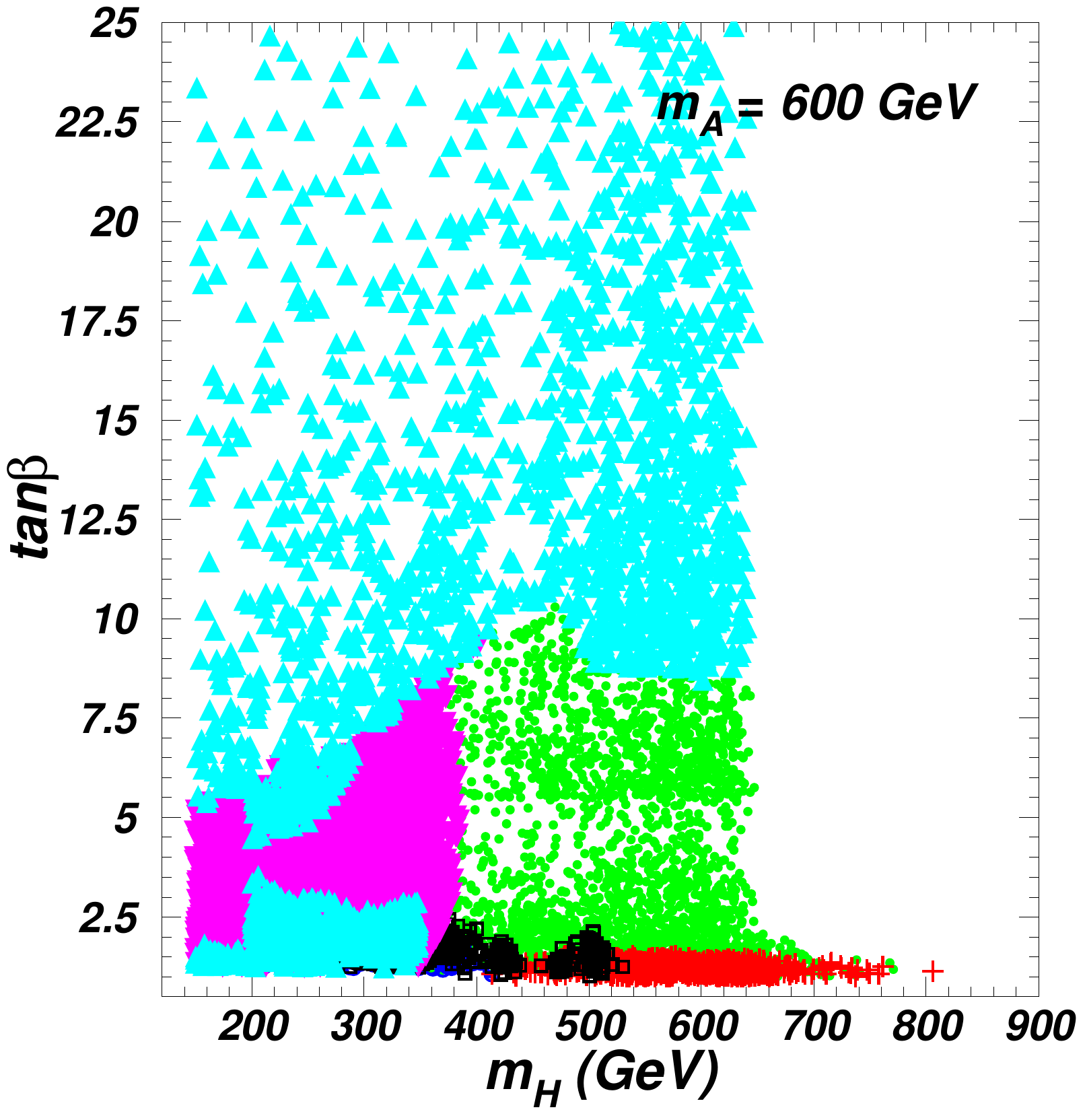,height=7cm}
 \epsfig{file=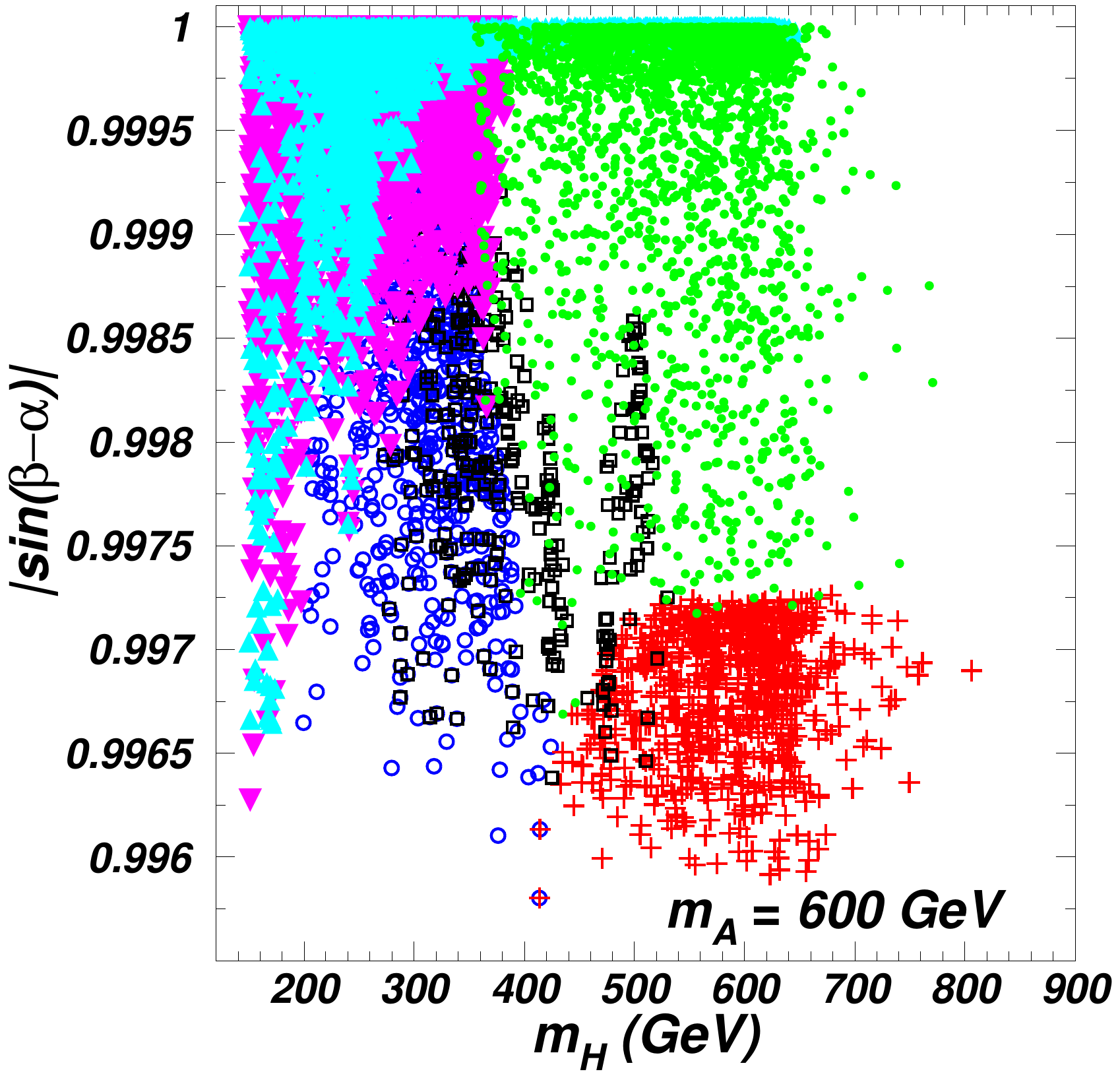,height=7cm}
\vspace{-0.2cm} \caption{The surviving samples with the
SM-like coupling projected on the planes of $m_H$ versus
$\tan\beta$ and $m_H$ versus $\mid\sin(\beta-\alpha)\mid$. All the samples
are allowed by the constraints of pre-LHC and the
125 GeV Higgs signal data. The triangles (sky blue), circles (royal blue), squares (black), inverted triangles (purple), 
and pluses (red) are respectively excluded by the $H/A\to \tau^+ \tau^-$, $H\to
WW,~ZZ,\gamma\gamma$, $H\to hh$, $A\to HZ$, and $A\to hZ$ channels at the LHC. The bullets (green) are
allowed by various LHC direct searches.} \label{lhc}
\end{figure}
In Fig. \ref{lhc}, we project the surviving samples with the SM-like
coupling on the planes of $m_H$ versus $\tan\beta$ and $m_H$
versus $\mid\sin(\beta-\alpha)\mid$ after imposing the constraints of
pre-LHC (denoting theoretical constraints, electroweak precision
data, the flavor observables, $R_b$, the exclusion limits from
searches for Higgs at LEP), the 125 GeV Higgs signal data, and the
searches for additional Higgses at the LHC. Note that in the region of $\sin(\beta-\alpha) < 0$, the signal data of the 125 GeV Higgs
require $\sin(\beta-\alpha)$ to nearly equal to -1.0, as shown in right panel of Fig. \ref{125higgs}. For such case,
the couplings of $H$ and $A$ are almost the same as those in the case of $\sin(\beta-\alpha)=1.0$. Therefore, we do not distinguish the sign
of $\sin(\beta-\alpha)$ when discussing the constraints on $m_H$ and $m_A$ from the LHC direct searches.

From Fig. \ref{lhc}, we find the joint constraints of
 $H/A\to\tau^+ \tau^-$, $A\to HZ$, $H\to WW,~ZZ,~\gamma\gamma$, and $H\to hh$ exclude the whole 
region of $m_H<360$ GeV.
The $H/A \to \tau^+ \tau^-$ channels impose upper bound on $\tan\beta$ in the whole range of $m_H$, and 
allow $m_H$ to vary from 150 GeV to 800 GeV for appropriate values of $\tan\beta$ and $\sin(\beta-\alpha)$.
The $A\to HZ$ channel does not constrain the parameter space of $m_H>$ 360 GeV since 
the branching ratio of $A\to HZ$ rapidly decreases with an increase of $m_H$.
 The limits of $A\to HZ$ channel
can be relaxed by a small $\mid\sin(\beta-\alpha)\mid$ which suppresses the $AHZ$ coupling.

The $H\to WW,~ZZ,~\gamma\gamma,~hh$ and $A\to hZ$ channels impose strong constraints on the regions with
small values of $\mid\sin(\beta-\alpha)\mid$ and $\tan\beta$ since the couplings of $HWW,~HZZ,~Hhh$ and $AhZ$
increase with decreasing of $\mid\sin(\beta-\alpha)\mid$, and $\sigma(gg\to H/A)$ is enhanced by the top quark loop 
for a small $\tan\beta$. In addition, the Fig. \ref{125higgs} shows that the 125 GeV Higgs signal data favor
a small $\tan\beta$ for a small $\mid\sin(\beta-\alpha)\mid$ in the case of the SM-like coupling.
With an increase of $m_H$, the $H\to t\bar{t}$ channel opens and enhances the total width of $H$ sizably, so that the
constraints from $H\to WW,~ZZ,~\gamma\gamma,~hh$ channels are relaxed.
Different from other channels, the $AhZ$ channel gives the constraints on the region with a large $m_H$.
This is because the width of $A\to HZ$ decreases with an increase of $m_H$, and thus $Br(A\to hZ)$ increases with $m_H$.

\section{The dark matter observables}
We use $\textsf{FeynRules}$ \cite{feyrule} to generate the model file, which is called by $\textsf{micrOMEGAs}$ \cite{micomega}
to calculate the relic density. In our scenario, the elastic scattering of $S$ on a nucleon receives
the contributions of the process with $t$-channel exchange of $H$, and
the spin-independent cross section between DM and nucleons is given by
\cite{sigis}
 \beq \sigma_{p(n)}=\frac{\mu_{p(n)}^{2}}{4\pi m_{S}^{2}}
    \left[f^{p(n)}\right]^{2},
\eeq where $\mu_{p(n)}=\frac{m_Sm_{p(n)}}{m_S+m_{p(n)}}$, \beq
f^{p(n)}=\sum_{q=u,d,s}f_{q}^{p(n)}\mathcal{C}_{S
q}\frac{m_{p(n)}}{m_{q}}+\frac{2}{27}f_{g}^{p(n)}\sum_{q=c,b,t}\mathcal{C}_{S
q}\frac{m_{p(n)}}{m_{q}},\label{fpn} \eeq with $\mathcal{C}_{S
q}= \frac{\lambda_H}{m_H^2} m_q y_H^q$. 
The values of the form factors $f_{q}^{p,n}$ and $f_{g}^{p,n}$ are extracted from $\textsf{micrOMEGAs}$ \cite{micomega}.

The Planck collaboration reported the relic density of cold DM in the universe,
 $\Omega_{c}h^2 = 0.1198 \pm 0.0015$ \cite{planck}. 
The XENON1T collaboration reported stringent upper bounds of 
the spin-independent DM-nucleon cross section \cite{xenon2018}. 
The Fermi-LAT searches for the DM annihilation from dwarf spheroidal satellite galaxies gave the upper limits on the averaged cross sections
of the DM annihilation to $e^+ e^-$, $\mu^+ \mu^-$, $\tau^+\tau^-$, $u\bar{u}$, $b\bar{b}$, and $WW$ \cite{fermi}.

\begin{figure}[tb]
 \epsfig{file=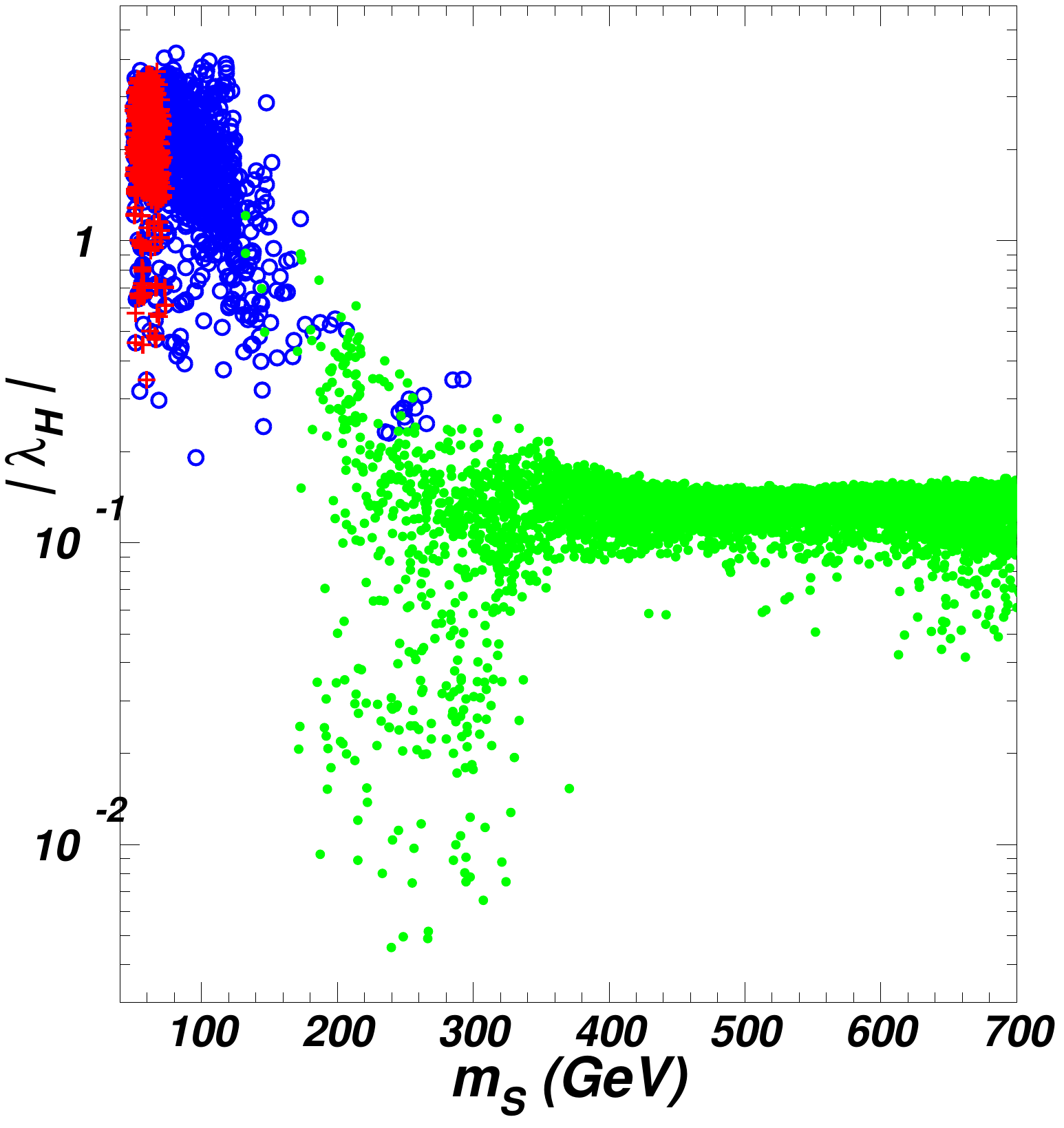,height=5.61cm}
\epsfig{file=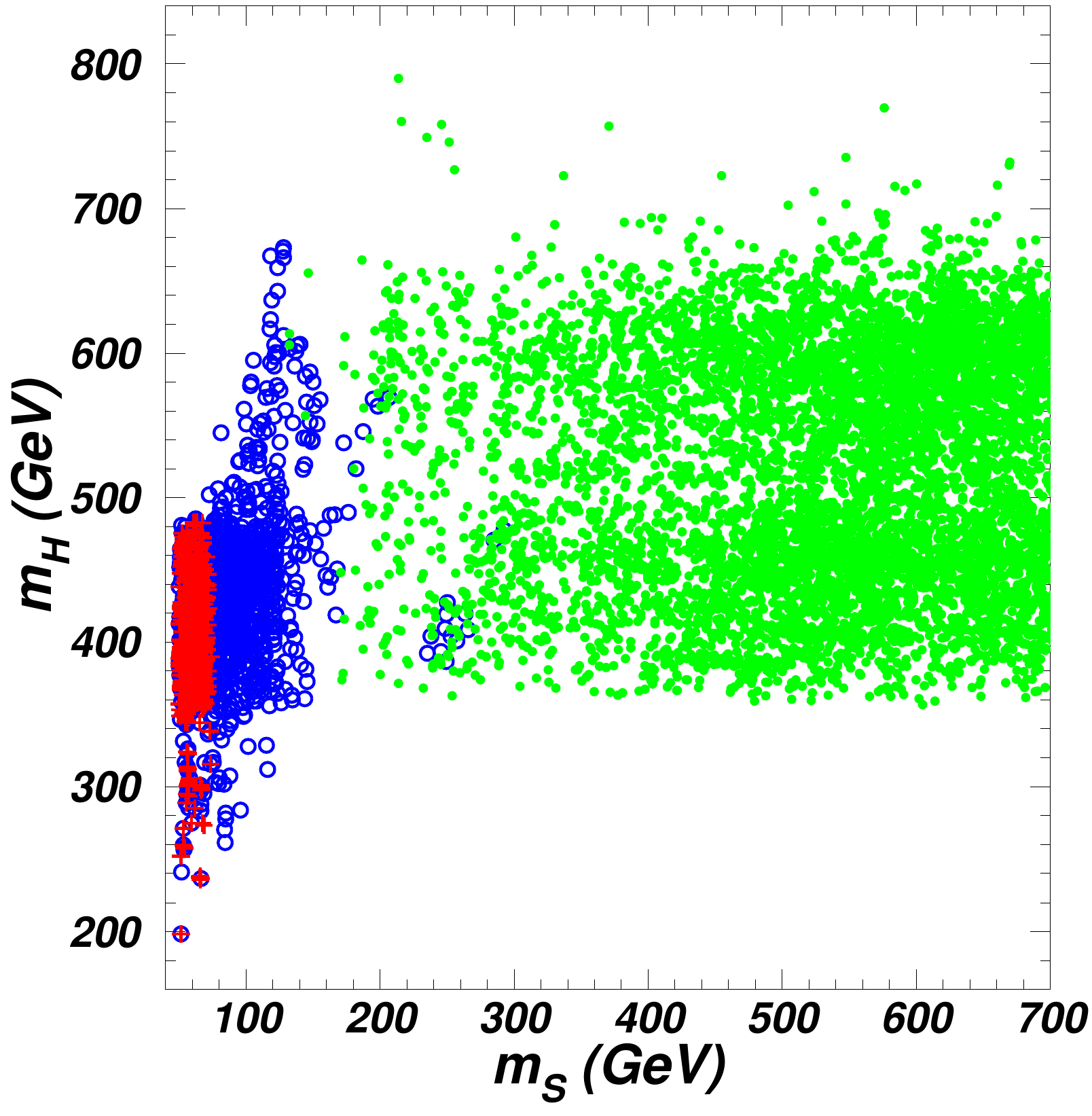,height=5.61cm}
\epsfig{file=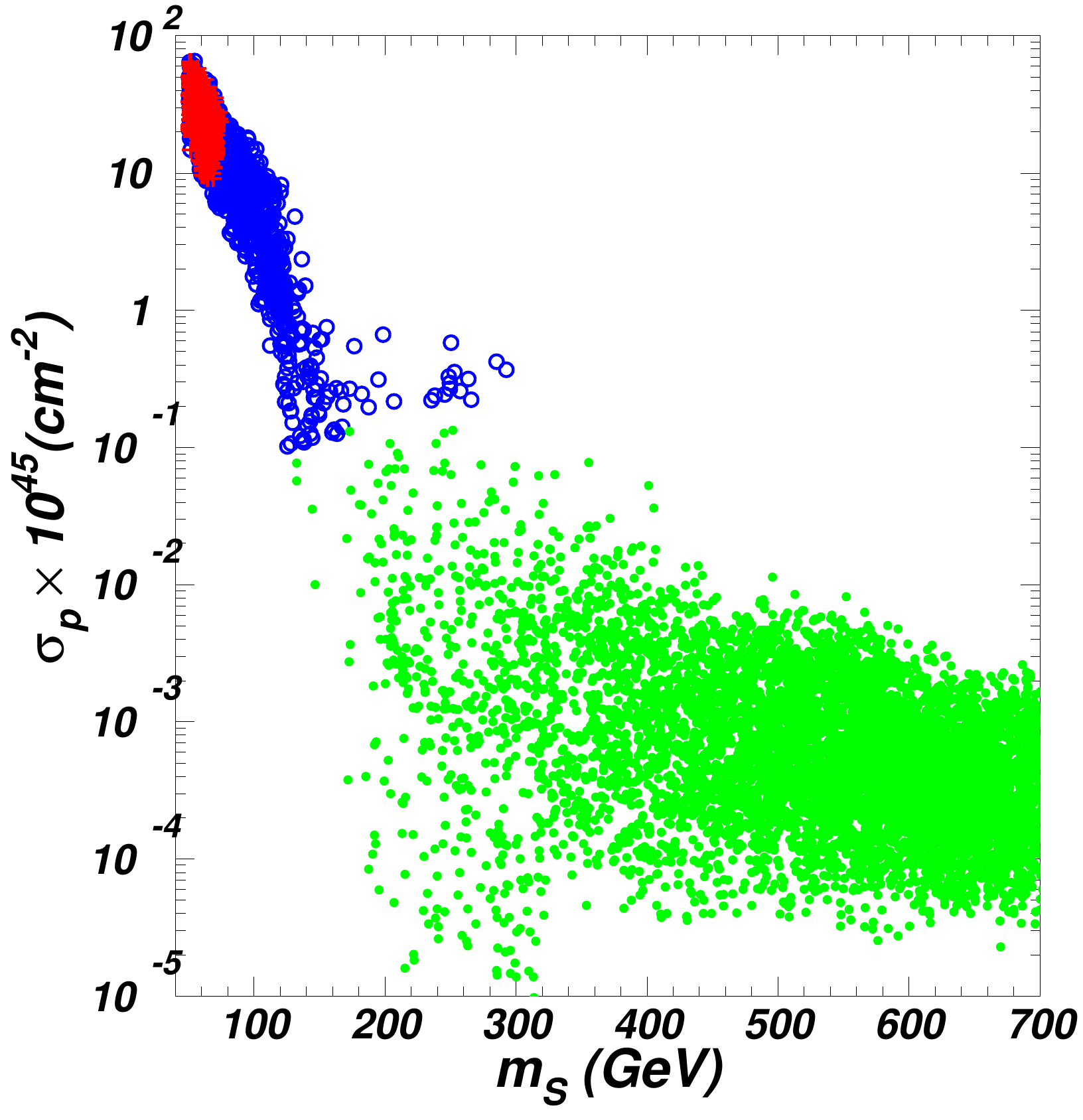,height=5.61cm}
\vspace{-0.5cm} \caption{The surviving samples projected on the
planes of $m_S$ versus $\lambda_H$, $m_S$ versus $m_H$, and $m_S$ versus $\sigma_{p}$.
All the samples are allowed by the constraints of "pre-LHC", the LHC Higgs data, and the relic density.
The circles (royal blue) and pluses (red) are respectively excluded by the experimental data of the XENON1T and Fermi-LAT, while
the bullets (green) are allowed.} \label{figxenon}
\end{figure}

In Fig. \ref{figxenon}, we project the surviving samples on the
planes of $\lambda_H$ versus $m_S$, $m_H$ versus $m_S$, and $\sigma_{p}$ versus $m_S$ after imposing the constraints of "pre-LHC",
 the Higgs data at the LHC, the relic density, XENON1T, and Fermi-LAT. The middle panel shows that the $H\to SS$ decay
weakens the constraints of the LHC Higgs data compared to Fig.\ref{lhc}. For example, $m_H$ is allowed to be as low as 200 GeV 
for a light DM. However, the upper bounds of the XENON1T and Fermi-LAT exclude $m_S < 130$ GeV and $m_H<$ 360 GeV.
In order to obtain the correct relic density, $\mid\lambda_H\mid$ is favored to increase with decreasing of $m_S$.
Thus, for a small $m_S$, a large $\mid\lambda_H\mid$ can enhance the spin-independent DM-nucleon cross section and the  
averaged cross sections of the today DM annihilation to the SM particles, leading that $m_S<130$ GeV and $m_S<$ 75 GeV
are respectively excluded by the experimental data of the XENON1T and Fermi-LAT.
For 180 GeV $<m_S<$ 340 GeV, $\mid\lambda_H\mid$ can be allowed to be smaller than 0.01 because of the resonant contribution at 
$2m_S\sim m_H$.

\section{Electroweak phase transition and gravitational wave}
The phase transition can proceed in
basically two different ways. In a first-order phase transition, at the critical temperature $T_C$, the two degenerate minima 
will be at different points in field space, typically with a potential barrier in between. For
a second order (cross-over) transition, the broken and symmetric minimum are not degenerate 
until they are at the same point in field space. In this paper we focus on the 
SFOEWPT, which is required by a successful explanation of the observed BAU
and can produce primordial GW signals.

\subsection{The thermal effective potential}
In order to examine electroweak phase transition (EWPT), we first take $h_1$, $h_2$, and $S_1$ as the field configurations, and 
obtain the field dependent masses of the scalars ($h,~H,~A,~H^{\pm},~S$), the Goldstone boson ($G,~G^{\pm}$), the gauge boson, and fermions.
The masses of scalars are given 
\begin{align}
\hm^2_{h,H,S} &=\rm{eigenvalues} ( \widehat{\mathcal{M}^2_P} ) \ , \\
\hm^2_{G,A} &=\rm{eigenvalues} ( \widehat{\mathcal{M}^2_A}) \ , \\
\hm^2_{G^\pm,H^\pm} &=\rm{eigenvalues}  (\widehat{\mathcal{M}^2_C})  \ ,
\end{align}
\begin{align}
\widehat{\mathcal{M}^2_P}_{11} &={3\lam_{1}\over 2} h^{2}_{1}+{\lam_{345} \over 2} h^{2}_2  + m^{2}_{12} \tb - {\lam_{1} \over 2} v^{2} \cb^2 - {\lam_{345} \over 2} v^{2} \sb^{2} + {\kappa_{1} \over 2} S_1^{2}\nonumber\\
\widehat{\mathcal{M}^2_P}_{22} &={3\lam_{2}\over 2} h^{2}_{2}+{\lam_{345} \over 2} h^{2}_1  + {m^{2}_{12} \over \tb} - {\lam_{2} \over 2} v^{2} \sb^2 - {\lam_{345} \over 2} v^{2} \cb^{2} + {\kappa_{2} \over 2} S_1^{2}\nonumber\\
\widehat{\mathcal{M}^2_P}_{33} &=m^2_{S}+ {\kappa_{1} \over 2} h^{2}_1+ {\kappa_{2} \over 2} h^{2}_2 + {\lam_S \over 2} S_1^{2} -{\kappa_{1} \over 2} v^{2} \cb^2- {\kappa_{2} \over 2} v^{2} \sb^2\nonumber\\
\widehat{\mathcal{M}^2_P}_{12} &=\widehat{\mathcal{M}^2_P}_{21}=\lam_{345} h_1 h_2 - m^{2}_{12}\nonumber\\
\widehat{\mathcal{M}^2_P}_{13} &=\widehat{\mathcal{M}^2_P}_{31}=\kappa_{1} h_1 S_1\nonumber\\
\widehat{\mathcal{M}^2_P}_{23} &=\widehat{\mathcal{M}^2_P}_{32}=\kappa_{2} h_2 S_1 \nonumber\\
\widehat{\mathcal{M}^2_A}_{11} &={\lam_{1}\over 2} h^{2}_{1}+ m^{2}_{12} \tb - {\lam_{1} \over 2} v^{2} \cb^2 - {\lam_{345} \over 2} v^{2} \sb^{2} +{(\lam_{3}+\lam_4-\lam_5) \over 2} h^{2}_2  + {\kappa_{1} \over 2} S_1^{2}\nonumber\\
\widehat{\mathcal{M}^2_A}_{22} &={\lam_{2}\over 2} h^{2}_{2}+ {m^{2}_{12} \over \tb} - {\lam_{2} \over 2} v^{2} \sb^2 - {\lam_{345} \over 2} v^{2} \cb^{2} +{(\lam_{3}+\lam_4-\lam_5) \over 2} h^{2}_1  + {\kappa_{2} \over 2} S_1^{2}\nonumber\\
\widehat{\mathcal{M}^2_A}_{12} &=\widehat{\mathcal{M}^2_A}_{21}=\lam_{5} h_1 h_2 - m^{2}_{12}\nonumber\\
\widehat{\mathcal{M}^2_C}_{11} &={\lam_{1}\over 2} h^{2}_{1}+ m^{2}_{12} \tb - {\lam_{1} \over 2} v^{2} \cb^2 - {\lam_{345} \over 2} v^{2} \sb^{2} +{\lam_{3} \over 2} h^{2}_2  + {\kappa_{1} \over 2} S_1^{2}\nonumber\\
\widehat{\mathcal{M}^2_C}_{22} &={\lam_{2}\over 2} h^{2}_{2}+ {m^{2}_{12} \over \tb} - {\lam_{2} \over 2} v^{2} \sb^2 - {\lam_{345} \over 2} v^{2} \cb^{2} +{\lam_{3} \over 2} h^{2}_1  + {\kappa_{2} \over 2} S_1^{2}\nonumber\\
\widehat{\mathcal{M}^2_C}_{12} &=\widehat{\mathcal{M}^2_C}_{21}={(\lam_{4}+\lam_{5}) \over 2} h_1 h_2 - m^{2}_{12},
\end{align}
where $\lambda_{345}=\lambda_3+\lambda_4+\lambda_5$, $c_\beta=\cos\beta$, and $s_\beta=\sin\beta$.

The masses of gauge boson are given
\begin{align}
\hm^2_{W^\pm} &= {1\over 4} g^2 \left(h^{2}_{1} +h^{2}_{2} \right),\nonumber\\
\hm^2_{Z} &= {1\over 4} (g^2+g'^2)  \left(h^{2}_{1} +h^{2}_{2} \right), \nonumber\\
\quad \hm^2_{\gamma}&=0. 
\end{align}

We neglect the contributions of light fermions, and only consider the masses of top quark and bottom quark,
\beq
\hm^2_t = {1\over 2} y^2_t h^{2}_{2}/{s_\beta^2},  ~~~~~
\hm^2_b = {1\over 2} y^2_b h^{2}_{1}/{c_\beta^2}. 
\eeq
where $y_t={\sqrt{2} m_t \over v}$ and $y_b={\sqrt{2} m_b \over v}.$

Now we study the effective potential with thermal correction. The thermal effective potential $V_{eff}$ in terms of the classical fields ($h_1,~h_2,~S_1$) is composed of four parts:
\begin{align}
V_{eff} (h_1,h_2,S_1,T)= &V_{0}(h_1,h_2,S_1) + V_{CW}(h_1,h_2,S_1) + V_{CT}(h_1,h_2,S_1) \nonumber\\
&+ V_{T}(h_1,h_2,S_1,T) + V_{ring}(h_1,h_2,S_1,T).
\label{veff0}
\end{align}
Where $V_{0}$ is the tree-level potential, $V_{CW}$ is the Coleman-Weinberg potential, 
$V_{CT}$ is the counter term, $V_{T}$ is the thermal correction, and $V_{ring}$ is the resummed daisy corrections.
In this paper, we calculate $V_{eff}$ in the Landau gauge.

We obtain the tree-level potential $V_0$ in terms of their classical fields ($h_1,~h_2,~S_1$) 
\begin{eqnarray} \label{v0} \mathcal{V}_{0} &=& 
\left[{1 \over 2} m_{12}^2 t_\beta - {1 \over 4}\lam_1 v^2 c_\beta^2- {1 \over 4}\lam_{345} v^2 s_\beta^2 \right]h_1^2\nonumber \\
&&+ \left[{1 \over 2} m_{12}^2 {1 \over t_\beta} - {1 \over 4}\lam_2 v^2 s_\beta^2- {1 \over 4}\lam_{345} v^2 c_\beta^2 \right]h_2^2\nonumber \\
&&+ {\lam_1 \over 8} h_1^4 + {\lam_2 \over 8} h_2^4 - m_{12}^2 h_1 h_2 + {1\over 4} \lam_{345} h_1^2 h_2^2\nonumber\\
&&+ {\kappa_1 \over 4} h_1^2 S_1^2 + {\kappa_2 \over 4} h_2^2 S_1^2 + {1 \over 2} m_S^2 S_1^2 +  {1 \over 24} \lambda_s S_1^4\nonumber\\
&&- {\kappa_1 \over 4} v^2 c_\beta^2 S_1^2 - {\kappa_2 \over 4} v^2 s_\beta^2 S_1^2.
\end{eqnarray}

The Coleman-Weinberg potential in the $\overline{\rm MS}$ scheme at 1-loop level has the form \cite{Coleman:1973jx}:
\beq
\label{eq:CWpot}
V_{\rm CW}(h_{1},h_2,S_1) = \sum_{i} (-1)^{2s_i} n_i\frac{\hm_i^4 (h_{1},h_2,S_1)}{64\pi^2}
\left[\ln \frac{\hm_i^2 (h_{1},h_2,S_1)}{Q^2}-C_i\right],
\end{equation}
where $i=h,H,A,H^\pm,S,G,G^\pm,W^\pm,Z,t,b$, and $s_i$ is the spin of particle i. $Q$ is a renormalization scale, and we take $Q^2=v^2$.
The constants $C_i =\frac{3}{2}$ for scalars or fermions and
$C_i = \frac{6}{5}$ for gauge bosons.
$n_i$ is the number of degree of freedom,
\begin{align}
&n_h=n_H=n_G=n_A=1,\nonumber\\ 
&n_{H^\pm}=n_{G^\pm}=2,\nonumber\\
&n_{W^\pm}=6,~n_{Z}=3,\nonumber\\
&n_{t}=n_{b}=12.
\end{align}

With $V_{CW}$ being included in the potential, the minimization conditions of scalar potential 
in Eq. (\ref{veff0}) and the CP-even mass matrix will be shifted
slightly. To maintain the minimization conditions at T=0, we add the so-called ``counter-terms" 
\begin{align}
\label{eq:Vct}
V_{\rm CT}&=\delta m_1^2 h_{1}^2+\delta m_2^2 h_{2}^2 +\delta \lam_1 h_{1}^4+\delta \lam_{12} h_{1}^2 h_{2}^2 +\delta \lam_2 h_{2}^4\nonumber\\
&+\delta m_0^2 S_1^2 + \delta\kappa_1 h_1^2 S_1^2 +\delta\kappa_2 h_2^2 S_1^2, 
\end{align}
where the relevant coefficients are determined by
\beq
\label{eq:V1der}
\frac{\partial V_{\rm CT}}{\partial h_{1}} = -\frac{\partial V_{\rm CW}}{\partial h_{1}}\;, \quad \frac{\partial V_{\rm CT}}{\partial h_{2}} = -\frac{\partial V_{\rm CW}}{\partial h_{2}},\; \quad \frac{\partial V_{\rm CT}}{\partial S_1} = -\frac{\partial V_{\rm CW}}{\partial S_1},
\eeq
\begin{align}
\label{eq:V2der}
\frac{\partial^{2} V_{\rm CT}}{\partial h_{1}\partial h_{1}} = - \frac{\partial^{2} V_{\rm CW}}{\partial h_{1}\partial h_{1}}\;, \quad
\frac{\partial^{2} V_{\rm CT}}{\partial h_{1}\partial h_{2}} = - \frac{\partial^{2} V_{\rm CW}}{\partial h_{1}\partial h_{2}}\;, \quad
\frac{\partial^{2} V_{\rm CT}}{\partial h_{2}\partial h_{2}} = - \frac{\partial^{2} V_{\rm CW}}{\partial h_{2}\partial h_{2}}\;, \nonumber\\
\frac{\partial^{2} V_{\rm CT}}{\partial S_1\partial S_1} = - \frac{\partial^{2} V_{\rm CW}}{\partial S_1\partial S_1}\;, \quad
\frac{\partial^{2} V_{\rm CT}}{\partial h_{1}\partial S_1} = - \frac{\partial^{2} V_{\rm CW}}{\partial h_{1}\partial S_1}\;, \quad
\frac{\partial^{2} V_{\rm CT}}{\partial h_{2}\partial S_1} = - \frac{\partial^{2} V_{\rm CW}}{\partial h_{2}\partial S_1}\;, 
\end{align}
which are evaluated at the EW minimum of $\{ h_{1}=v\cb, h_{2}=v\sb, S_1=0 \}$ on both sides.
As a result, the VEVs of $h_{1}$, $h_{2}$, $S_1$ and the CP-even mass matrix will not be shifted.

It is a well-known problem that the second derivative of the
Coleman-Weinberg potential at $T=0$ suffers from logarithmic divergences originating
from the vanishing Goldstone masses. To solve the divergence problem,
we take a straightforward approach of imposing an
IR cut-off at $m^2_{IR} = m^2_h$ for the masses of Goldstone boson of the divergent terms, 
which gives a good approximation to the exact procedure of
on-shell renormalization, as argued in \cite{PT_2HDM1.5}.

The thermal contributions $V_T$ to the potential can be written as \cite{v1t}
\beq
\label{potVth}
 V_{\rm th}(h_{1},h_{2},S_1,T) = \frac{T^4}{2\pi^2}\, \sum_i n_i J_{B,F}\left( \frac{ \hm_i^2(h_{1},h_{2},S_1)}{T^2}\right)\;,
\eeq
where $i=h,H,A,H^\pm,S,G,G^\pm,W^\pm,Z,t,b$, and the functions $J_{B,F}$ are 
\beq
\label{eq:jfunc}
J_{B,F}(y) = \pm \int_0^\infty\, dx\, x^2\, \ln\left[1\mp {\rm exp}\left(-\sqrt{x^2+y}\right)\right].
\eeq

Finally, the thermal corrections with resumed ring diagrams are given \cite{vdai1,vdai2}
\beq
V_{\rm ring}\left(h_{1},h_{2},S_1, T\right) =-\frac{T}{12\pi }\sum_{i} n_{i}\left[ \left( \bar{M}_{i}^{2}\left(h_{1},h_{2},S_1,T\right) \right)^{\frac{3}{2}}-\left( \hm_{i}^{2}\left(h_{1},h_{2},S_1,T\right) \right)^{\frac{3}{2}}\right] ,
\label{eq:daisy}
\eeq
where $i=h,H,A,H^\pm,S,G,G^\pm,W^\pm_L,Z_L,\gamma_L$. The $W^\pm_L,~Z_L$, and $\gamma_L$ are the longitudinal gauge bosons with
$n_{W^\pm_L}=2,~n_{Z_L}=n_{\gamma_L}=1$.
The thermal Debye masses $\bar{M}_{i}^{2}\left(h_{1},h_2,S_1,T\right)$ are the eigenvalues of the full mass matrix, 
\begin{equation}
\label{eq:thermalmass}
\bar{M}_{i}^{2}\left( h_{1},h_{2},T\right) ={\rm eigenvalues} \left[\widehat{\mathcal{M}_{X}^2}\left( h_{1},h_{2}\right) +\Pi _{X}(T)\right]  ,
\end{equation}%
where $X=P,A,C$. $\Pi_X$ are given by 
\begin{align}
\Pi_{P11} &= \left[{9g^2\over 2} + {3g'^2\over 2} + {6y_b^2 \over \cb^2} + 6\lam_{1} +4\lam_{3} +2\lam_4 + \kappa_{1} \right] {T^2 \over 24}\nonumber\\
\Pi_{P22} &= \left[{9g^2\over 2} + {3g'^2\over 2} + {6y_t^2 \over \sb^2} + 6\lam_{2} +4\lam_{3} +2\lam_4 + \kappa_{2} \right] {T^2 \over 24}\nonumber\\
\Pi_{P33} &= \left[4\kappa_{1} +4\kappa_{2} +\lam_S \right] {T^2 \over 24}\nonumber\\
\Pi_{P13} &=\Pi_{P31} =\Pi_{P23} =\Pi_{P32}=0\nonumber\\
\Pi_{A11} &= \Pi_{C11}=\Pi_{P11}\nonumber\\
\Pi_{A22} &= \Pi_{C22}=\Pi_{P22}\nonumber\\
\Pi_{A12} &= \Pi_{A21}=\Pi_{C12} = \Pi_{C21}=0.
\end{align}

The physical mass of the longitudinally polarized $W$ boson is 
\beq
\bar{M}_{W^{\pm}_L}^2 = {1 \over 4} g^2 (h^2_1+h^2_2) + 2 g^2 T^2.
\eeq 
The physical mass of the longitudinally polarized $Z$ and $\gamma$ boson
\beq
\bar{M}_{Z_L,\gamma_L}^2 = \frac{1}{8} (g^2+g'^2) (h^2_1+h^2_2) + (g^2 + g^{\prime 2} )T^2 \pm \Delta, 
\eeq
with 
\beq
\Delta^2 =\frac{1}{64}  (g^2 + g^{\prime 2} )^2(h_{1}^2 + h_{2}^2+8T^2)^2
- g^2 g^{\prime 2} T^2 ( h_{1}^2 + h_{2}^2 + 4 T^2). 
\eeq

\subsection{Calculation of electroweak phase transition and gravitational wave}
In a first-order cosmological phase transition, bubbles nucleate and expand, converting
the high-temperature phase into the low-temperature one.
The bubble nucleation rate per unit volume at finite temperature is given by \cite{bubble-0,bubble-1,bubble-2}
\begin{eqnarray}
\Gamma \ \approx \ A(T)e^{-S_E(T)},
\end{eqnarray}
where $A(T)\sim T^4$ is a prefactor and $S_E$ is the Euclidean action
\begin{eqnarray}
S_E(T)=\frac{S_3(T)}{T}  = \ \int dx^3\bigg[\frac{1}{2}\big(\frac{d \phi}{dr}\big)^2+V(\phi,T)\bigg].
\end{eqnarray}
 At the nucleation temperature $T_n$, the thermal tunneling probability for bubble nucleation per horizon volume and per horizon time is 
of order one, and the conventional condition is $\frac{S_3(T)}{T}\approx 140$. The bubbles nucleated within one Hubble patch proceed to expand and collide, until the
entire volume is filled with the true vacuum.

There are two key parameters characterizing the dynamics of the EWPT, 
$\beta$ and $\alpha$. $\beta$ describes roughly the inverse time duration of the strong first order phase transition,
\begin{eqnarray}
\frac{\beta}{H_n}=T\frac{d (S_3(T)/T)}{d T}|_{T=T_n}\; ,
\end{eqnarray}
where $H_n$ is the Hubble parameter at the bubble nucleation temperature $T_n$.
$\alpha$ is defined as the vacuum energy released from the phase transition normalized by the total radiation energy
density $\rho_R$ at $T_n$,
      \begin{eqnarray}
      \alpha=\frac{\Delta\rho}{\rho_R}=\frac{\Delta\rho}{\pi^2 g_{\ast} T_n^4/30}\;,
      \end{eqnarray}
where $g_{\ast}$ is the effective number of relativistic degrees of freedom.
We use the numerical package CosmoTransitions \cite{cosmopt} and PhaseTracer \cite{Athron:2020sbe} to analyze the phase transition and 
computes quantities related to cosmological phase transition.

In a radiation-dominated Universe, there are three sources of GW production at a EWPT: bubble collisions, in which the localized energy density generates a quadrupole contribution to the stress-energy tensor, which in turn gives rise to GW, plus sound waves in the plasma and magnetohydrodynamic (MHD) turbulence. The total resultant energy
density spectrum can be approximately given as,
\begin{equation}
\Omega_{\text{GW}}h^{2} \ \simeq \ \Omega_{\rm col}h^{2}+\Omega_{\rm sw}h^{2}+\Omega_{\rm turb}h^{2}.
\end{equation}
Recent studies show that the energy deposited in the bubble walls is negligible, despite the
possibility that the bubble walls can run away in some circumstances \cite{gw-coll-1}. Therefore, although a
bubble wall can reach relativistic speed, its contribution to GW can generally be
neglected \cite{gw-coll-2,gw-coll-3}. Therefore, in the following discussions we do not include the contribution from bubble collision $\Omega_{\rm col}$.

The GW spectrum from the the sound waves can be obtained by fitting to the result of
numerical simulations \cite{gw-sw},
\begin{eqnarray}
\Omega_{\textrm{sw}}h^{2} & \ = \ &
2.65\times10^{-6}\left( \frac{H_{n}}{\beta}\right)\left(\frac{\kappa_{v} \alpha}{1+\alpha} \right)^{2}
\left( \frac{100}{g_{\ast}}\right)^{1/3} v_{w}\nonumber \\
&&\times  \left(\frac{f}{f_{sw}} \right)^{3} \left( \frac{7}{4+3(f/f_{\textrm{sw}})^{2}} \right) ^{7/2} \ ,
\label{eq:soundwaves}
\end{eqnarray}
where $f_{\text{sw}}$ is the present peak frequency of the spectrum,
\begin{equation}
f_{\textrm{sw}} \ = \ 
1.9\times10^{-5}\frac{1}{v_{w}}\left(\frac{\beta}{H_{n}} \right) \left( \frac{T_{n}}{100\textrm{GeV}} \right) \left( \frac{g_{\ast}}{100}\right)^{1/6} \textrm{Hz} \,.
\label{fsw}
\end{equation}
$v_w$ is the wall velocity, and the factor $\kappa_{v}$ is the fraction of latent heat transformed into the kinetic energy of the fluid.
 $\kappa_{v}$ and $v_w$ are difficult to compute, and involves certain assumptions regarding the
dynamics of the bubble walls. On the other hand, successful electroweak
baryogenesis scenarios favor lower wall velocity $v_w\leq$ $0.15 - 0.3$ \cite{ewbg-vw},
 which allows the effective diffuse of particle asymmetries near
the bubble wall front. In Ref. \cite{new-ewbg-vw}, however, it is pointed out that the relevant velocity for electroweak
baryogenesis is not really $v_w$,
 but the relative velocity between the bubble wall and the plasma in the deflagration front. As a result, the electroweak
baryogenesis is not 
necessarily impossible even in the case with large $v_w$.
Therefore, in this paper we take two different cases of $v_w$ and $\kappa_{v}$ \cite{1004.4187}:
\begin{itemize}
  \item For small wall velocity:  $v_w=0.3$ and 
\beq
\kappa_v \simeq v_w^{6/5} 
\frac{6.9 \alpha}{1.36 - 0.037 \sqrt{\alpha} + \alpha}\ .
\eeq

\item For very large wall velocity: $v_w=0.9$ and
\beq
\kappa_v \simeq \frac{\alpha}
{0.73 + 0.083 \sqrt{\alpha} + \alpha}\ .
\eeq
\end{itemize}

Considering Kolmogorov-type turbulence as proposed in
Ref. \cite{mhd-type}, the GW spectrum from the MHD turbulence has the form \cite{mhd-1,mhd-2},
\begin{eqnarray}
\Omega_{\textrm{turb}}h^{2} & \ = \ & 
3.35\times10^{-4}\left( \frac{H_{n}}{\beta}\right)\left(\frac{\kappa_{turb} \alpha}{1+\alpha} \right)^{3/2} \left( \frac{100}{g_{\ast}}\right)^{1/3} v_{w} \nonumber \\
  && \times \frac{(f/f_{\textrm{turb}})^{3}}
  {[1+(f/f_{\textrm{turb}})]^{11/3}(1+8\pi f/h_{n})},
\label{eq:mhd}
\end{eqnarray}
with the red-shifted Hubble rate at GW generation
\beq
h_n=1.65\times 10^{-5}\left(\frac{T_n}{100\textrm{GeV}}\right) \left(\frac{g_{\ast}}{100}\right)^{\frac{1}{6}}\textrm{Hz}.
\eeq
 The peak frequency $f_{\rm turb}$ is given by
\begin{equation}
f_{\textrm{turb}} \ = \ 2.7\times10^{-5}\frac{1}{v_{w}}\left(\frac{\beta}{H_{n}} \right) \left( \frac{T_{n}}{100\textrm{GeV}} \right) \left( \frac{g_{\ast}}{100}\right)^{1/6} \textrm{Hz} \,.
\end{equation}
The energy fraction transferred to the MHD turbulence $\kappa_{\text{turb}}$ can vary
between $5\%$ to $10\%$ of $\kappa_v$~\cite{gw-sw}. Here we take $\kappa_{\text{turb}} = 0.1 \kappa_v$.

For both sound wave and turbulence
contribution as shown in Eq. (\ref{fsw}) and Eq. (\ref{eq:mhd}), the amplitudes of the GW spectra are proportional to
$v_w$ and the peak frequencies shift as $1/v_w$. Therefore, one changes in the wall velocity approximately
have an order one effect on the spectrum and peak frequencies. 

\subsection{Results and discussions}
\begin{figure}
 \epsfig{file=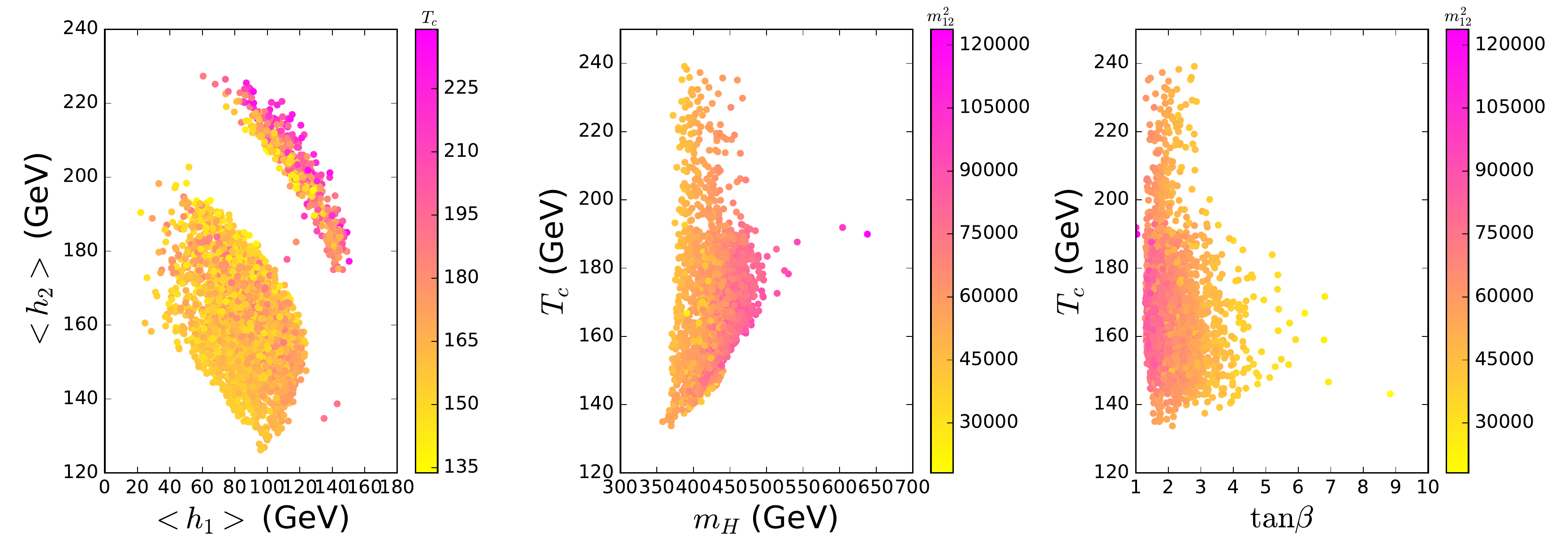,height=6.2cm}
\vspace{-1.0cm} \caption{The surviving samples projected on the
planes of $<h_1>$ versus $<h_2>$,
$T_c$ versus $m_H$, and $T_c$ versus $\tan\beta$. All the samples achieve a SFOEWPT.} \label{tcmh}
\end{figure}

\begin{figure}[tb]
 \epsfig{file=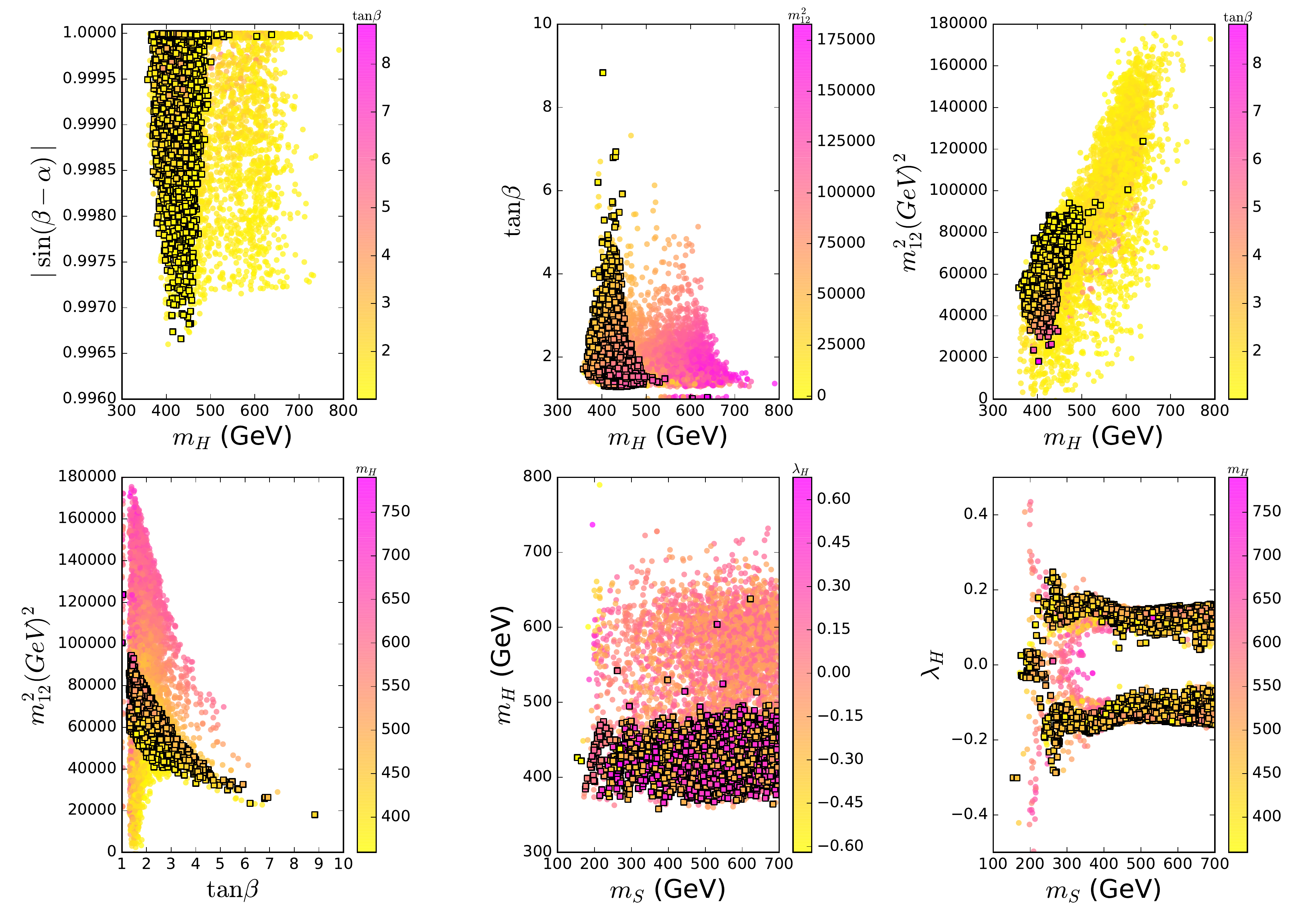,height=12cm}
\vspace{-1.1cm} \caption{The surviving samples projected on the
planes of $\mid\sin(\beta-\alpha)$, $\tan\beta$, $m_{12}^2$ versus $m_H$, $m_{12}^2$ versus $\tan\beta$, and $m_S$ versus $m_H$, $\lambda_H$.
All the samples are allowed by the constraints of "pre-LHC", the LHC Higgs data, and the DM observables.
The squares achieve a SFOEWPT, and bullets fail.} \label{ptmh}
\end{figure}
The strength of the electroweak phase transition is quantified as
\beq
\xi_c=\frac{v_c}{T_c}
\eeq
with $v_c=\sqrt{<h_1>^2+<h_2>^2}$ at critical temperature $T_c$. The global minimum of potential has $<A> = 0$
because of the CP-conserving case.
In order to avoid washing out the baryon number
generated during the phase transition, a SFOEWPT is required and the conventional condition is $\xi_c\geq 1$.

After imposing the constraints of "pre-LHC",
 the LHC Higgs data, the relic density, XENON1T, and Fermi-LAT, we scan over the parameter space in the previous selected scenario.
We find some surviving samples which can achieve a SFOEWPT, and these samples are projected in Fig. \ref{tcmh} and Fig. \ref{ptmh}.
For all the surviving samples, at $T_c$ the two degenerate minima of potential are respectively at ($<h_1>,~<h_2>,~0$) and (0,~0,~0).
In the process of EWPT, $<S_1>$ always has no VEV. 

From Fig. \ref{tcmh}, we find that $<h_1>$ and $<h_2>$ can vary in the ranges of 20 GeV $\sim$ 150 GeV and 125 GeV $\sim$ 230 GeV 
with $T_c$ varying from 134 GeV to 240 GeV. $T_c$ tends to increase with $m_H$, and has a relative small value for a large 
$\tan\beta$. It should also be noted that the relic abundance of the
DM is achieved by the thermal freeze-out in the early universe when the
temperature was about $T\sim m_S/25$. In the model, $T_c$ is much larger than $m_S/25$ for 50 GeV $<m_S<700$ GeV.
Therefore, the EWPT hardly affects the thermal freeze-out process of DM.

From Fig. \ref{ptmh} , we find that a SFOEWPT favors a small $m_H$, namely a large mass splitting between $m_H$ and $m_A$, which is consistent with
Refs. \cite{PT_2HDM2,PT_2HDM3}. Most of samples lie in the region of $m_H<500$ GeV, and there are several samples with $m_H>$ 500 GeV when 
$\mid \sin(\beta-\alpha)\mid$ is very closed to 1.0. Also a SFOEWPT favors $m_{12}^2$ to increase with $m_{H}$ and decrease with an increase of $\tan\beta$. There is a relative strong correlation
between $m_{12}^2$ and $\tan\beta$, and $m_{12}^2$ is imposed upper and lower bounds for a given $\tan\beta$.
 With an increase of $\tan\beta$, $m_{12}^2$ is stringently restricted by the theoretical constraints and the LHC Higgs data, leading that it 
is difficult to achieve a SFOEWPT. Thus, most of samples lie in the
region of small $\tan\beta$. The requirement of a SFOEWPT is not sensitive to $m_S$, and disfavors $\mid\lambda_H\mid >$ 0.3.

\begin{figure}[tb]
 \epsfig{file=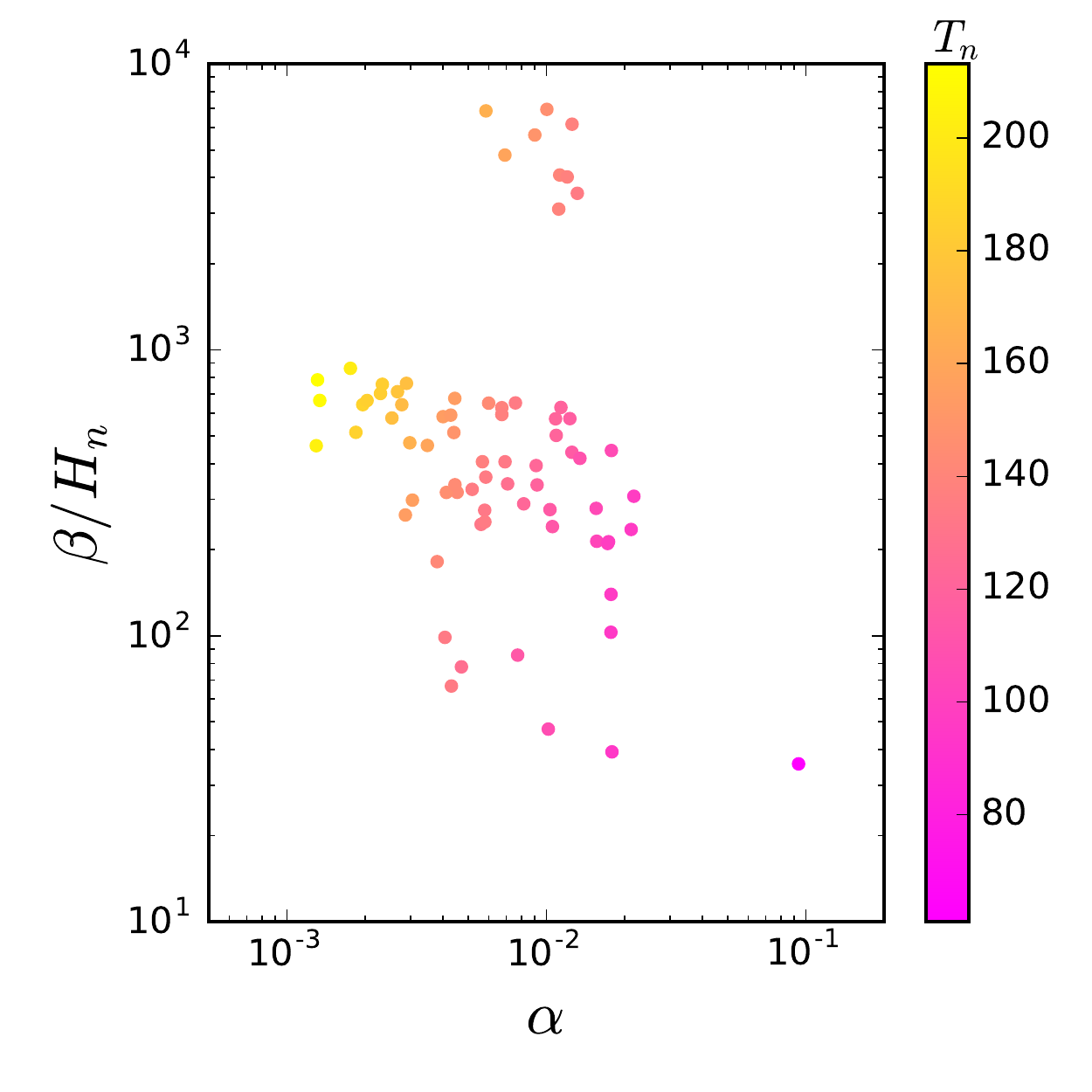,height=8cm}
\vspace{-0.9cm} \caption{The parameters $\alpha$ and $\beta/H_n$ characterizing the dynamics of the SFOEWPT.} \label{betaalpha}
\end{figure}

Now we examine two key parameters $\alpha$ and $\beta/H_n$ which characterize the dynamics of the SFOEWPT, and
govern the strength of GW spectra. A larger $\alpha$ and smaller $\beta/H_n$ can lead to stronger GW signals.
In addition to the conditions of the successful bubble nucleations, we require 
\beq
\xi_n=\frac{v_n}{T_n} \geq  1
\eeq
with $v_n=\sqrt{<h_1>^2+<h_2>^2}$ at the nucleation temperature $T_n$.
In fact, this is a more precise condition of SFOEWPT than $\xi_c\geq 1$.
Also note that
there generically exists a difficulty for solving bounce solution in a very thin-walled bubble, including
the package CosmoTransitions \cite{BubbleProfiler}. Therefore, we will neglect the samples with very thin-walled bubble.
Consider the constraints discussed above, we find some surviving samples, and the corresponding $\alpha$ and $\beta/H_n$
are shown in Fig. \ref{betaalpha}.  

\begin{table}
\begin{footnotesize}
\begin{tabular}{| c | c | c | c |}
\hline
 & ~~~~\textbf{BP1}~~~~ &~~~~  \textbf{BP2}~~~~ \\
\hline
$\sin(\beta-\alpha)$ & 0.9998  & 0.9991 \\
$\tan\beta$          & 1.95    & 1.87\\
$m_H$ (GeV)          & 369.55  & 387.97\\
$m_{H^\pm}$ (GeV)    & 620.8   & 618.31\\
$m_{12}^2$ (GeV$)^2$ & 53049.1 & 53649.1\\
$m_S$ (GeV)          & 479.2   & 501.7\\
$\lambda_H$          & 0.133   & -0.129\\
$\lambda_S$          & 12.3    & 10.93\\
$T_c$ (GeV)          & 135.7   & 160.0\\
$T_n$ (GeV)          & 61.0    & 95.0\\
$\beta/H_n$          & 35.6    & 102.8\\
$\alpha$             & 0.094   & 0.018\\
\hline
\end{tabular}
\end{footnotesize}
\caption{Input and output parameters for two benchmark points for fixed $m_h$=125 GeV, $m_A$=600 GeV and $\lambda_h=0$.}
\label{tabgrav}
\end{table}

\begin{figure}[tb]
 \epsfig{file=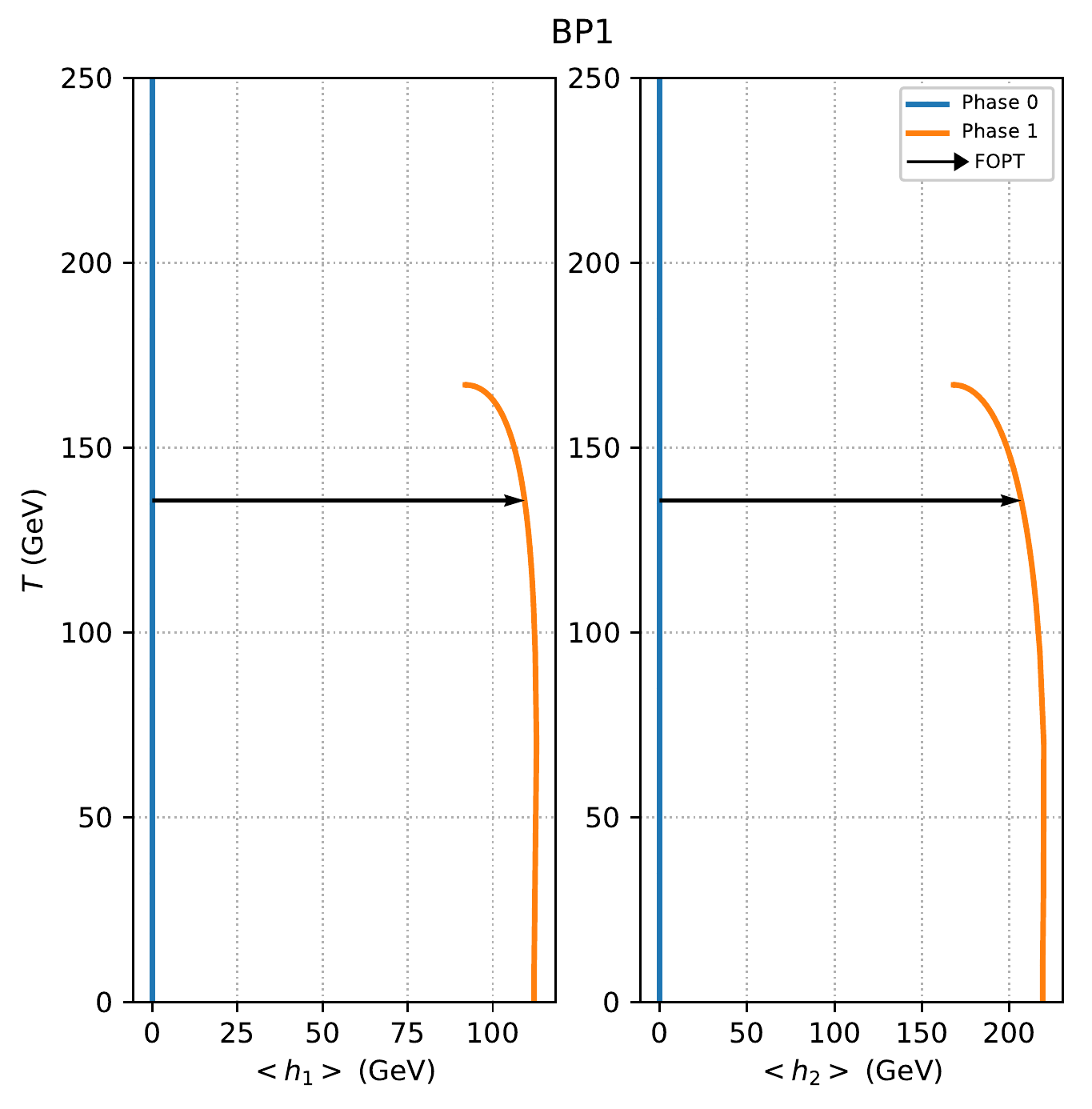,height=8cm}
 \epsfig{file=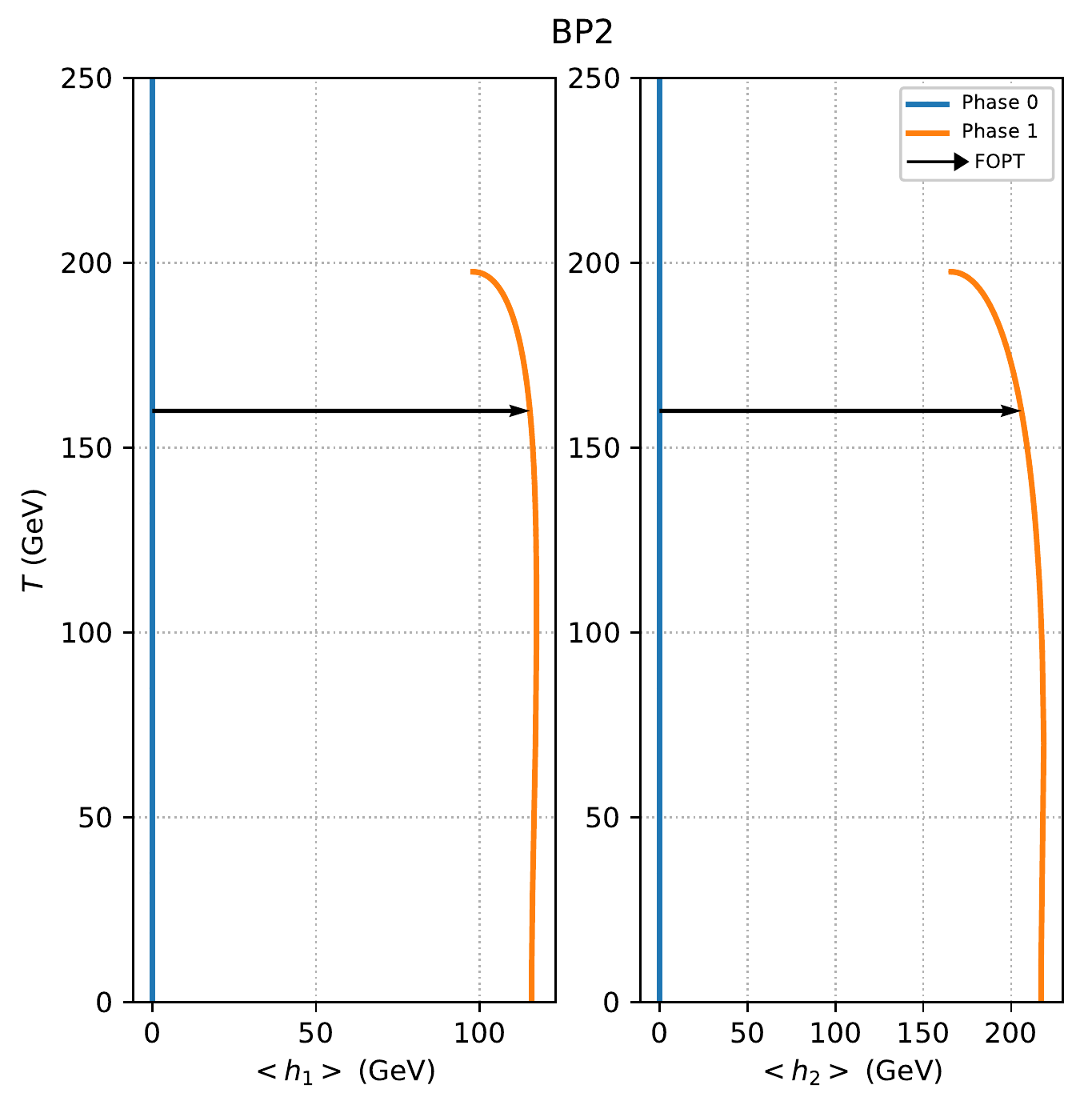,height=8cm}
\vspace{-0.2cm} \caption{Phase structures for BP1~(left) and BP2~(right). The lines show the field configurations at a particular minimum as a function of temperature. The arrows indicate that at that temperature ($T_C$) the two phases linked by the arrows are degenerate, and can
achieve the first order phase transition (FOPT).} \label{figphase}
\end{figure}

\begin{figure}[tb]
 \epsfig{file=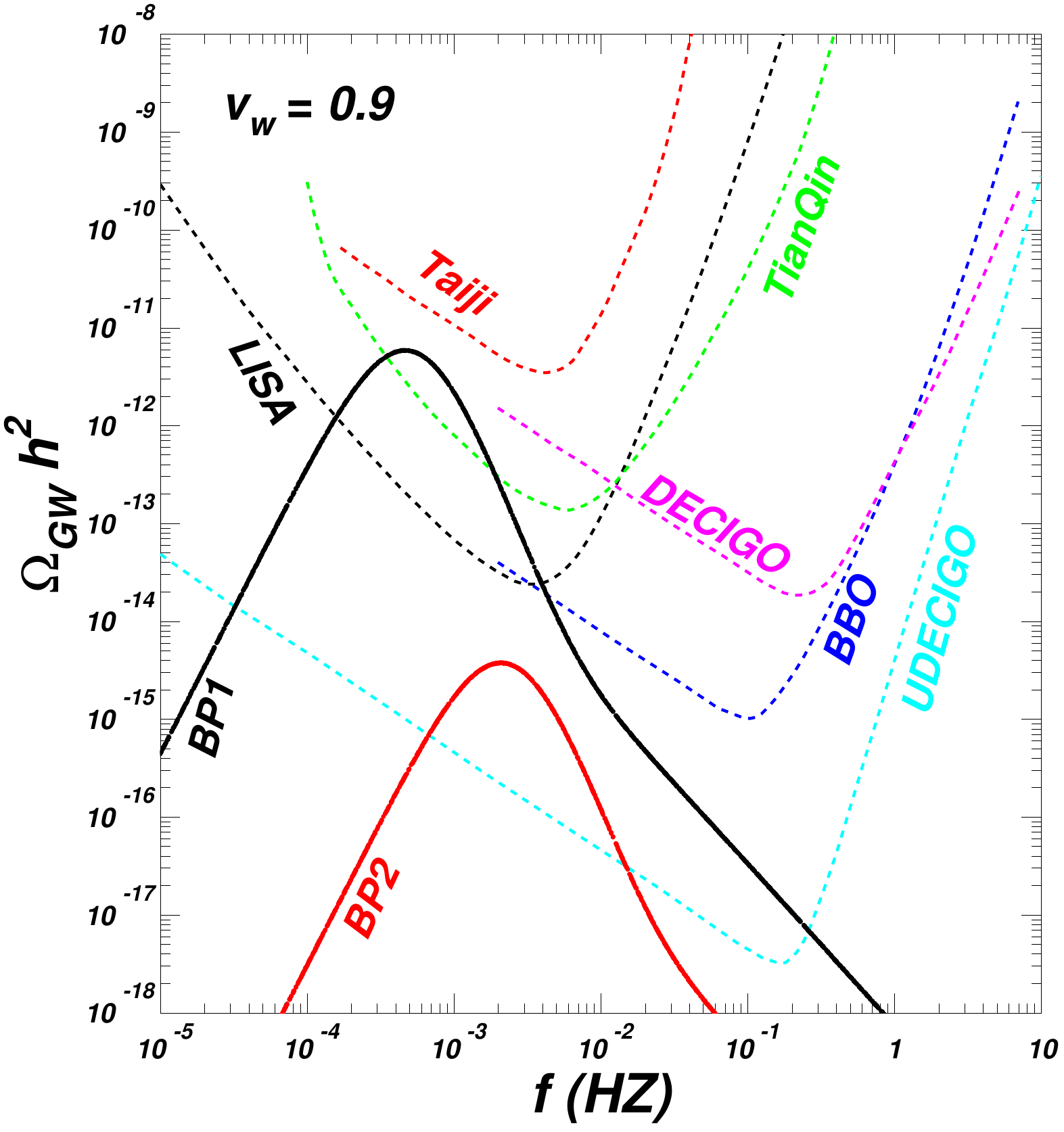,height=8cm}
 \epsfig{file=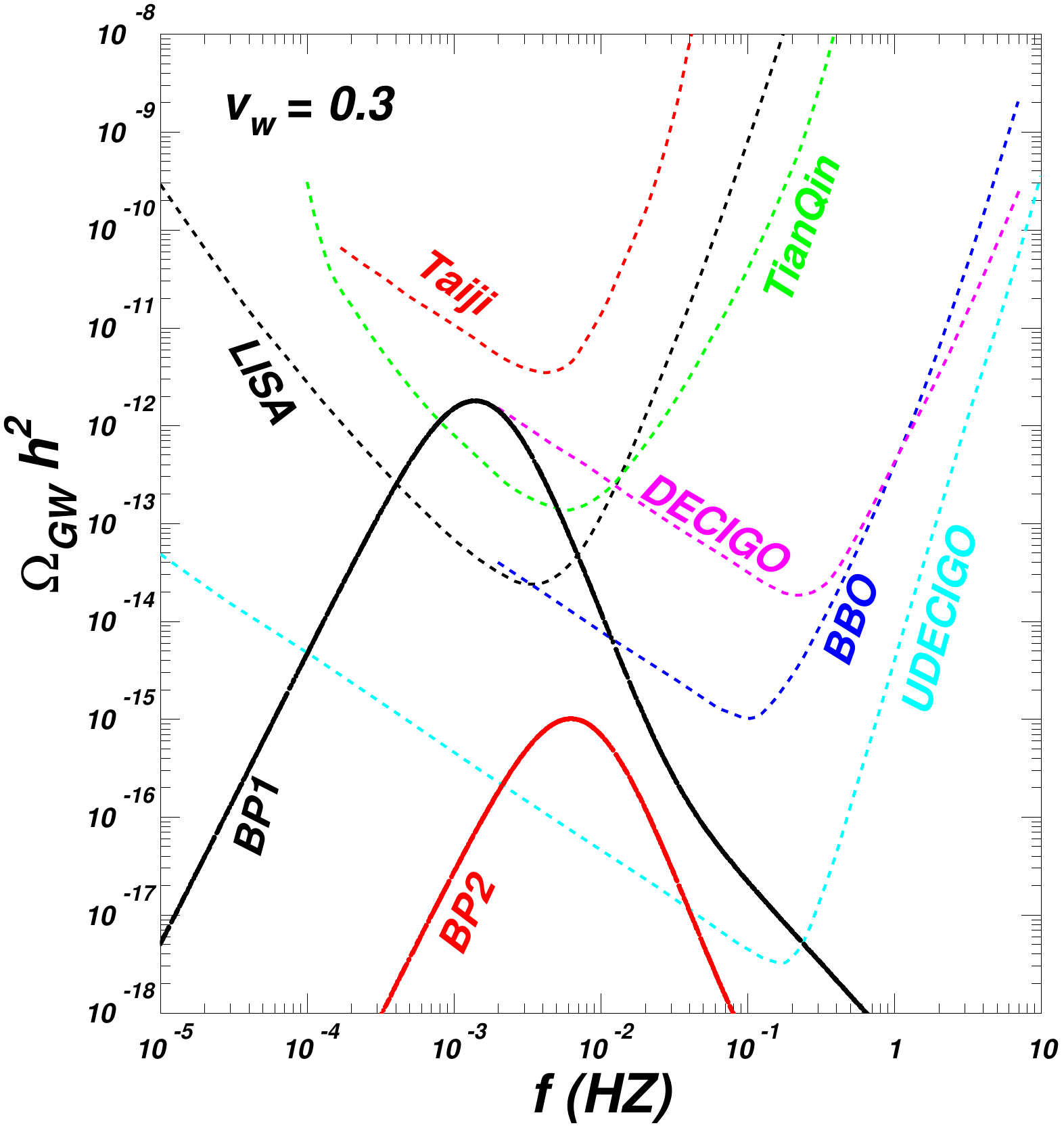,height=8cm}
\vspace{-0.2cm} \caption{Gravitational wave spectra for BP1 and BP2.} \label{figgrav}
\end{figure}

The $\beta/H_n$ may characterize the inverse time duration of the EWPT. A small $\beta/H_n$ means a long EWPT, and gives strong GW signals.
For the GW coming from the sound waves in the plasma, the GW signal will continue being generated and the energy density of the
GW is thus proportional to the duration of the EWPT if the mean square fluid velocity of the
plasma is non-negligible \cite{gw-sw}. In addition, a large $\beta/H_n$ can enhance the peak frequency of the GW spectra.
The parameter $\alpha$ describes the amount of energy released during the EWPT, and therefore a large $\alpha$ leads to strong GW signals.

We pick out two benchmark points (BPs), and examine the corresponding GW spectra.
Table \ref{tabgrav} shows the input and output parameters of the BPs. 
 Their phase histories are exhibited in Fig.~\ref{figphase} on filed configurations versus temperature plane. The filed configuration $S_1$ is not shown as the minima at any temperatures locate at $<S_1>=0$. 
In Fig. \ref{figgrav}, we show predicted GW spectra for our BPs along with expected sensitivities of various future
interferometer experiments, and find that the amplitudes of the GW spectra reach the sensitivities of LISA, TianQin, BBO, DECIGO, UDECIGO for BP1 (UDECIGO for BP2).

\section{Conclusion}
We examine the status of the 2HDMIID confronted with 
the recent LHC Higgs data, the DM observables and SFOEWPT, and 
discuss the detectability of GW at the future GW detectors.
We choose the heavy CP-even Higgs $H$ as the only portal between the DM and
SM sectors, and focus on the case of the 125 GeV Higgs with the SM-like coupling.
We find that for $m_A=600$ GeV, $m_S<$ 130 GeV and $m_H<$ 360 GeV are excluded by the joint constraints of the 
125 GeV Higgs signal data, the searches for additional Higgs via $H/A\to\tau^+ \tau^-$, $A\to HZ$, $H\to WW,~ZZ,~\gamma\gamma,~hh$ at the LHC as well as the relic density, XENON1T.

A SFOEWPT can be achieved in the many regions of $m_H<500$ GeV and $m_A=600$ GeV, favors a small $\tan\beta$, and is not 
sensitive to the mass of DM. We find the benchmark points for which the predicted GW spectra can reach
 the sensitivities of LISA, TianQin, BBO, DECIGO, and UDECIGO.

\section*{Acknowledgment}
We thank L. Bian, Wei Chao and Huai-Ke Guo for helpful discussions.
This work was supported by the National Natural Science Foundation
of China under grant 11975013, by the Natural Science Foundation of
Shandong province (ZR2017MA004 and ZR2017JL002), and by the ARC Centre of Excellence for Particle Physics at the Tera-scale under the grant CE110001004. This work
is also supported by the Project of Shandong Province Higher Educational Science and
Technology Program under Grants No. 2019KJJ007.

\end{document}